\documentclass[aoas,MSNbibl,nameyear,seceqn,dvips]{arximspdf}
\usepackage{graphicx}

\doi{10.1214/11-AOAS524}
\volume{6}
\issue{2}
\pubyear{2012}
\firstpage{669}
\lastpage{696}

\makeatletter
\newcommand\bxi{\boldsymbol{\xi}}
\newcommand\bfeta{\boldsymbol{\eta}}
\newcommand\bSig{{\boldsymbol{\Sigma}}}
\newcommand\bA{\mathbf{A}}
\newcommand\bC{\mathbf{C}}
\newcommand\bh{\mathbf{h}}
\newcommand\bR{\mathbf{R}}
\newcommand\bS{\mathbf{S}}
\newcommand\bU{\mathbf{U}}
\newcommand\bV{\mathbf{V}}
\newcommand\bw{\mathbf{w}}
\newcommand\bx{\mathbf{x}}
\newcommand\by{\mathbf{y}}
\newcommand\bZ{\mathbf{Z}}
\newcommand\sg{{\sigma}}
\newcommand\eg{{\epsilon}}
\renewcommand{\epsilon}{\varepsilon}
\newcommand\ag{{\alpha}}
\newcommand\la{{\lambda}}
\newcommand\kg{{\kappa}}
\newcommand\ga{{\gamma}}
\newcommand\dg{{\delta}}
\newcommand\cov{{\operatorname{Cov}}}
\newcommand{\nn}{\nonumber}
\newcommand\cS{{\mathcal S}}
\newcommand\hla{{\hat\la}}
\newcommand\hv{{\hat v}}
\newcommand\hu{{\hat u}}
\newcommand\ip{{i^\prime}}
\newcommand\jp{j^\prime}
\makeatother

\begin{document}
\begin{frontmatter}

\title{Estimation and testing for
spatially indexed curves with application
to ionospheric and magnetic~field~trends\thanksref{T1}}
\runtitle{Spatially indexed curves and ionospheric trends}
\thankstext{T1}{Supported
in part by NSF Grants DMS-08-04165 and DMS-09-31948.}

\begin{aug}
\author[A]{\fnms{Oleksandr} \snm{Gromenko}\ead[label=e1]{agromenko@gmail.com}},
\author[A]{\fnms{Piotr} \snm{Kokoszka}\corref{}\ead[label=e2]{Piotr.Kokoszka@usu.edu}},
\author[B]{\fnms{Lie} \snm{Zhu}\ead[label=e3]{zhu@cc.usu.edu}}
\and
\author[B]{\fnms{Jan}~\snm{Sojka}\ead[label=e4]{sojka@cc.usu.edu}}
\runauthor{Gromenko, Kokoszka, Zhu and Sojka}
\affiliation{Utah State University}
\address[A]{O. Gromenko\\
P. Kokoszka\\
Department of Mathematics and Statistics\\
Utah State University\\
Logan, Utah 84322-3900\\
USA\\
\printead{e1}\\
\hphantom{\textsc{E-mail}: }\printead*{e2}} %adresu isvedimo komanda
%gale!
\address[B]{L. Zhu\\
J. Sojka\\
Department of Physics and \\
\quad Center for
Atmospheric and Space Science\\
Utah State University\\
Logan, Utah 84322-4405\\
USA\\
\printead{e3}\\
\hphantom{\textsc{E-mail}: }\printead*{e4}}
\end{aug}

% HISTORY:
\received{\smonth{1} \syear{2011}}
\revised{\smonth{8} \syear{2011}}

% ABSTRACT
%
\begin{abstract}
We develop methodology for the estimation of the
functional mean and the functional principal components when the
functions form a spatial process. The data consist of curves
$X(\mathbf{s}_k; t),  t \in[0, T],$ observed at spatial locations $\mathbf{s}_1,
\mathbf{s}_2, \ldots, \mathbf{s}_N$. We propose several methods, and
evaluate them
by means of a simulation study.
Next, we develop a significance test for the correlation of two such
functional spatial fields. After validating the finite sample
performance of this test by means of a simulation study, we apply it
to determine if there is correlation between long-term trends in the
so-called critical ionospheric frequency and decadal changes in the
direction of the internal magnetic field of the Earth. The test
provides conclusive evidence for correlation, thus solving a long-standing space physics conjecture. This conclusion is not apparent
if the spatial dependence of the curves is neglected.
\end{abstract}

% KEYWORDS
%

\end{frontmatter}

%s1 ###
\section{Introduction} \label{si}
The contribution of this paper to
statistics is two-fold: (1)~we develop estimation methodology for the
functional mean and the functional principal components (FPCs) when
the functions form a spatial field; (2) we propose a significance test
to determine if two families of curves observed at the same spatial
locations are uncorrelated. The contribution to space physics
consists in solving a controversy regarding the impact of long-term
changes in the internal magnetic field of the Earth on long-term
ionospheric trends. The required physics background is provided
later in this section, and in Section~\ref{sapp-te}.

The data is modeled as curves $X(\mathbf{s}_k; t),  t \in[0,
T],$ observed at spatial locations $\mathbf{s}_1, \mathbf{s}_2, \ldots, \mathbf{s}_N$. Such
functional data structures are quite common, but typically the spatial
dependence and the spatial distribution of the points~$\mathbf{s}_k$ are
not\vadjust{\goodbreak}
taken into account. A~fundamental question is how to estimate the
mean function of curves indexed by spatial locations. Clearly, curves
located at close-by points look similar and must be given smaller
weights than curves at points far apart.
In addition to the mean function, FPCs play a~fundamental role in
functional data analysis.
Good estimators of FPCs
are needed to construct reliable testing and classification
procedures, but such issues have been addressed only in the contexts
of independent curves, with focus on sparsity and measurement error.
The geophysical data that motivate this research are
available at fine temporal grids and are measured with errors that are
negligible relative to the objectives of the statistical analysis. A~focus of recent geophysical research is on the detection and
estimation of global and/or regional long-term trends (the global
warming paradigm), so before a~statistical analysis is undertaken, the
data are typically smoothed to remove daily or even annual
periodicity. The question we address is how to combine the temporal
trajectories available at many spatial locations to obtain meaningful
summary trends. We argue that one can do better than using simple
averaging. The focus of this paper is thus on combining information
from spatially dependent curves, which are smooth and available at all
time points.

%f1 ###
\begin{figure}[b]

\includegraphics{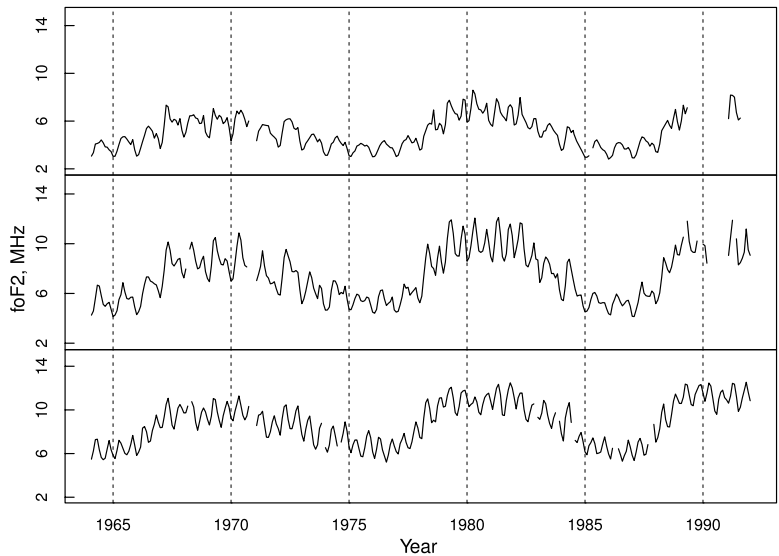}

\caption{F2-layer critical frequency curves at three locations.
Top to bottom (latitude in parentheses): Yakutsk (62.0),
Yamagawa (31.2), Manila (14.7). The functions exhibit a latitudinal trend
in amplitude.}
\label{ffoF2}
\end{figure}
%f2 ###
\begin{figure}

\includegraphics{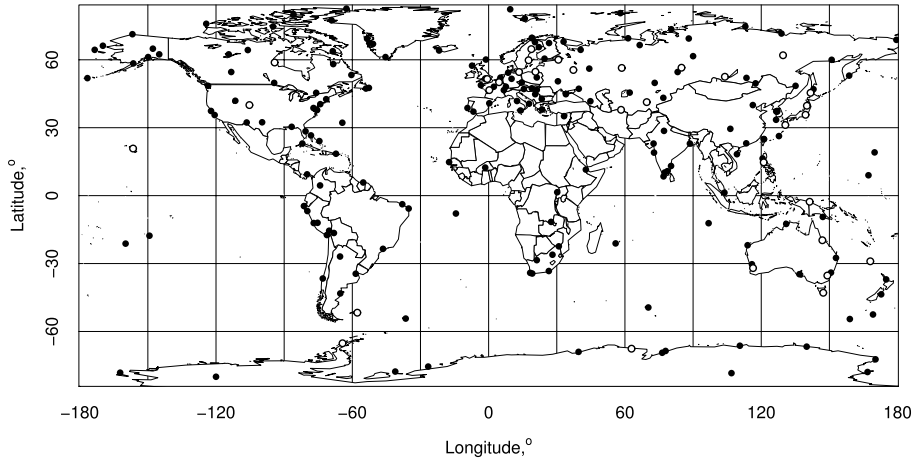}

\caption{Locations of 218 ionosonde stations.
Circles represent the 32 stations with the longest complete records.}
\label{floc}
\end{figure}

\begin{figure}

\includegraphics{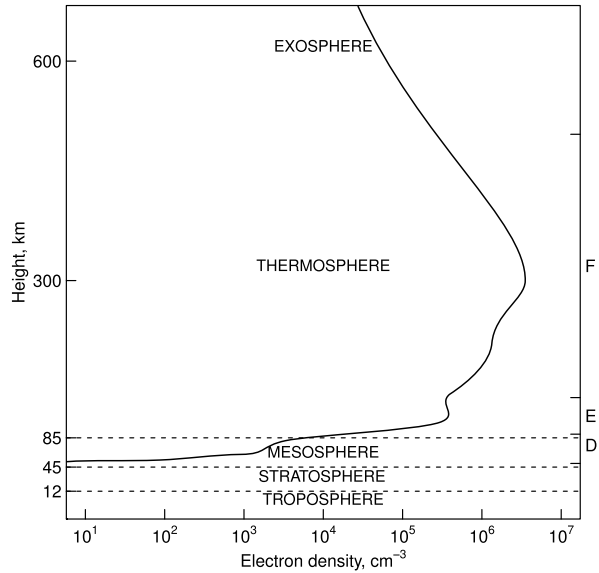}

\caption{Typical profile of day time ionosphere.
The curve shows electron density as a function of height.
The right vertical axis indicates the D, E and F regions.}
\label{fai}
\end{figure}

Many environmental and geophysical data sets fall into the
framework considered in this paper.
The data set that motivated this research consists of the curves of
the ionospheric F2-layer critical frequency, foF2. Three such curves
are shown in Figure \ref{ffoF2}. In principle, foF2 curves are
available at over 200 locations throughout the globe (see Figure
\ref{floc}), but sufficiently
complete data are available at only 30--40 locations which are very
unevenly spread; for example, there is a dense network of
observatories over Europe and practically no data over the oceans.
The study of this data set has been motivated by the hypothesis of
Roble and Dickinson (\citeyear{roble-dickinson-1989}), who suggested
that the increasing
amounts of (radiative) greenhouse gases should lead to global cooling
in mesosphere and thermosphere, as opposed to the global warming in
lower troposphere; cf. Figure \ref{fai}. Rishbeth (\citeyear{rishbeth-1990})
pointed out that such cooling would result in a thermal contraction
and the global lowering of the ionospheric peak densities, which can
be computed from the critical frequency foF2. The last twenty years
have seen very extensive research in this area; see
La\v{s}tovi\v{c}ka et al. (\citeyear
{lastovicka-2008})
for a partial overview. One of the
difficulties in determining a global trend is that the foF2 curves
appear to exhibit trends in opposing directions over various regions.
A possible explanation suggests that these trends are caused by
long-term trends in the magnetic field of the Earth. There is, however,
currently not agreement in the space physics community if this is
indeed the case. In general, to make any trends believable, a suitable
statistical modeling and a proper treatment of ``errors and
uncertainties'' is called for
[Ulich, Clilverd and Rishbeth (\citeyear
{ulich-clilverd-risbeth-2003})]. This paper makes a
contribution in this direction. Space physics data measured at
terrestrial observatories always come in the form of temporal curves
at fixed spatial locations. In Maslova et al. (\citeyear
{maslova-kokoszka-s-z-2009}),
Maslova et al. (\citeyear{maslova-kokoszka-s-z-2010}) and Maslova et al.
(\citeyear{maslova-kokoszka-s-z-2010PSS}) the tools of functional
data analysis were used to study such data, but the spatial dependence
of the
curves was not fully exploited.

%f3 ###

Spatio-temporal modeling has received a great deal of attention
of late; see Part~V of Gelfand et al. (\citeyear{gelfand-etal-2010}) and
Chapters 3, 4 and 6 of Finkenstaedt, Held and Isham (\citeyear
{finkenstaedt-etal-2007})
which discuss spatio-temporal models for geostatistical data.
There has, however, not been much research specifically on
spatially indexed functional data; Delicado et al. (\citeyear
{delicado-2010}) review recent
contributions. For geostatistical functional data, several approaches to
kriging have been proposed; see Yamanishi and Tanaka (\citeyear
{yamanishi-tanaka-2003}),
Nerini, Monestiez and Mant\'e (\citeyear{nerini-monestiez-mantea-2010}),
Giraldo, Delicado and Mateu (\citeyear{giraldo-delicado-mateu-2011a})
and Bel et al. (\citeyear{bel-2011}).

Throughout the paper, $\{ X(\mathbf{s})\}$ denotes a random field defined on
a spatial domain and taking values in the Hilbert space
$L^2=L^2([0,1])$ with the inner product
\[
\langle f, g \rangle= \int_0^1 f(t) g(t)\,dt,    \qquad  f,g \in L^2.
\]
The value of the function $X(\mathbf{s})\in L^2$ at time $t\in[0,1]$
is denoted by $X(\mathbf{s}; t)$.
We postulate the model
%
%e1.1 ###
\begin{equation} \label{emodel}
X(\mathbf{s}; t) = \mu(t) + \sum_{i=1}^\infty
\xi_{i}(\mathbf{s}) e_{i}(t),  \qquad
\xi_{i}(\mathbf{s}) = \langle X(\mathbf{s}) - \mu, e_i \rangle,
\end{equation}
where the $e_i$ form a complete orthonormal system.
Note that the mean
function $\mu$ and the FPCs $e_i$ do not depend on $\mathbf{s}$.
A sufficient condition for this is that the distribution in $L^2$
of the function $X(\mathbf{s})$ does not depend on the location $\mathbf{s}$.
A stronger sufficient condition is the strict stationarity
of the field $\{ X(\mathbf{s})\}$.

For the applications we have in mind, it is enough to assume that the
spatial domain is a subset of the plane or a two-dimensional sphere.
On the plane, the distance between points is the usual Euclidean
distance; on the sphere, we use the chordal distance defined as the
Euclidean distance in the three-dimensional space. The reason for
using the chordal distance is that any spatial covariance functions in
$\mathbb{R}^3$ restricted to the unit sphere is then also a covariance
function on the sphere. Denoting the latitude by $L$ and the
longitude by $l$, the chordal distance, $0 \leq d_{k,\ell}\leq2$,
between two points, $\mathbf{s}_k, \mathbf{s}_\ell$, on the unit sphere is
given by
%
%e1.2 ###
\begin{equation}\qquad
\label{edistance}
d_{k,\ell} = 2 \biggl[\sin^2 \biggl(\frac{L_k -
L_{\ell}}{2} \biggr)+ \cos{L_{k}}\cos{L_{\ell}}\sin^2
\biggl(\frac{l_k - l_{\ell}}{2} \biggr) \biggr]^{1/2}.
\end{equation}

For arbitrary (not necessarily spatially indexed) functions,
$X_1, X_2, \ldots, X_N$, the sample mean is defined as
$
\bar X_N = N^{-1} \sum_{n=1}^N X_n,
$
and the sample covariance operator as
\[
\widehat C(x) = N^{-1} \sum_{n=1}^N
 [ \langle(X_n - \bar X_N), x \rangle(X_n - \bar X_N) ],
  \qquad  x \in L^2.
\]
The sample FPCs are computed
as the eigenfunctions of $\widehat C$. These are the estimates
produced by several software packages, including the
popular \texttt{R} package \texttt{fda}; see Ramsay, Hooker and Graves
(\citeyear{ramsay-hooker-graves-2009}).
The consistency of the sample mean
and the sample FPCs relies on the assumption that the functional
observations form a simple random sample. If the functions $X_k =
X(\mathbf{s}_k)$
are spatially distributed, the sample mean and the sample FPCs need
not even be consistent; see H\"orman and Kokoszka (\citeyear
{hormann-kokoszka-2011}).
This happens
if the spatial dependence is strong or if there are clusters of the points
$\mathbf{s}_k$. We will demonstrate that better estimators are available and
we will use them as part of the
procedure for testing the independence of two functional
fields $\{ X(\mathbf{s}), \mathbf{s}\in\bS\}$ and $\{ Y(\mathbf{s}), \mathbf{s}\in\bS\}$.
The procedure is based on the observed pairs of functions
$(X(\mathbf{s}_k), Y(\mathbf{s}_k)),  1 \le k \le N$. The test we propose
is applied to ionosonde (X) and magnetic (Y) curves, and
conclusively shows that the temporal evolution of these two families
is strongly correlated.

The remainder of the paper is organized as follows. Sections
\ref{smean} and \ref{sPC} focus, respectively, on the estimation of
the mean function and the FPCs in a spatial setting.
Section~\ref{sfs} demonstrates by means of a simulation study that
the methods we propose improve on the standard approach, and discusses
their relative performance and computational cost. In
Section~\ref{ste} we develop a test for the correlation of two
functional spatial fields. This test requires estimation of a
covariance tensor. After addressing this issue in
Section~\ref{sestSig}, we study in Section~\ref{ssp-cor} the
finite sample properties of several implementations of the test.
Finally, in Section~\ref{sapp-te} we apply the methodology developed
in the previous section to test for the correlation between the
ionospheric critical frequency and magnetic curves.

%s2 ###
\section{Estimation of the
mean function} \label{smean}
We propose three methods of estimating
the mean function $\mu$, which we call M1, M2, M3. As will become
apparent in this section, several further variants, not discussed here,
are conceivable. But the results of Section~\ref{sfs}
show that while all these methods offer an improvement over
the simple sample mean, their performance is comparable.
We represent the observed functions as
%
%e2.1 ###
\begin{equation} \label{e1p}
X(\mathbf{s}_k; t) = \mu(t) + \eg(\mathbf{s}_k; t),    \qquad  k =1,2,\ldots, N,
\end{equation}
where $\eg$ is an unobservable field with $E\eg(\mathbf{s}; t) = 0$.
All methods
assume that the function valued field $\eg(\cdot)$ is strictly stationary
and isotropic, even though weaker, more technical assumptions could be
made for the specific methods. Methods M1 and M2 are akin to the
kriging technique advocated by Giraldo, Delicado and Mateu (\citeyear
{giraldo-delicado-mateu-2011a})
in that they treat the curves as single entities and seek to minimize
the integrated mean squared error. Method M3 is similar in spirit
to the approach to functional kriging developed by
Nerini, Monestiez and Mant\'e (\citeyear
{nerini-monestiez-mantea-2010}) and
Giraldo, Delicado and Mateu (\citeyear{giraldo-delicado-mateu-2011b})
who use cokriging of basis coefficients.

Methods M1 and M2 estimate $\mu$ by the weighted average
%
%e2.2 ###
\begin{equation} \label{ehat-mu}
\hat\mu_N = \sum_{n=1}^N w_n X(\mathbf{s}_n).
\end{equation}
The optimal weights $w_k$ are defined to minimize
$
E \Vert\sum_{n=1}^N w_n X(\mathbf{s}_n) -\mu\Vert^2
$
subject to the condition $ \sum_{n=1}^N w_n = 1$
($\| x \|^2 = \int_0^1 x^2(t)\,dt$).
Using the method of the Lagrange multiplier, we see
that the unknowns $w_1, w_2, \ldots, w_N, r$ are solutions
to the system of $N+1$ equations
%
%e2.3 ###
\begin{equation} \label{ew-mean}
\sum_{n=1}^N w_n =1,     \qquad
\sum_{k=1}^N w_k C_{kn} -r = 0,    \qquad  n=1,2, \ldots, N,
\end{equation}
where
%
%e2.4 ###
\begin{equation} \label{eCkl}
C_{k\ell} = E [\langle\eg(\mathbf{s}_k), \eg(\mathbf{s}_\ell)\rangle].
\end{equation}
Set $\bw=(w_{1},\ldots,w_{N})^{T}$.
An easy way to solve the equations in (\ref{ew-mean}) is to compute
$\mathbf{v}=\bC^{-1} {\mathbf1}$, where $\bC= [C_{k\ell}, 1 \le k,
\ell\le N]$,
and then set $\bw= a\mathbf{v}$, where $a$ is a constant such that
${\mathbf1}^{T} \bw=1$.

\textit{Method} M1. At each time point $t_j$, we fit a parametric spatial model
to the scalar field $X(\mathbf{s}; t_j)$.
To focus attention, we provide formulas\vadjust{\goodbreak} for the exponential model
%
%e2.5 ###
\begin{equation} \label{emod-M1}
\cov(X(\mathbf{s}_k; t_j), X(\mathbf{s}_\ell; t_j))
= \sg^2(t_j) \exp \biggl\{ - \frac{d(\mathbf{s}_k, \mathbf{s}_\ell)}{\rho
(t_j)}  \biggr\}.
\end{equation}
It is clear how they can be modified for other popular models.

Observe that under model (\ref{emod-M1}),
\begin{eqnarray*}
C_{k\ell}
&=& E \int \bigl( X(\mathbf{s}_k; t) - \mu(t)  \bigr) \bigl ( X(\mathbf{s}_\ell; t) -\mu(t)  \bigr)\,dt\\
&=& \int\cov(X(\mathbf{s}_k; t_j), X(\mathbf{s}_\ell; t_j))\,dt\\
& =& \int\sg^2(t) \exp \biggl\{ - \frac{d(\mathbf{s}_k, \mathbf{s}_\ell
)}{\rho(t)}  \biggr\}\,dt.
\end{eqnarray*}

One way to estimate $C_{k\ell}$ is to set
%
%e2.6 ###
\begin{equation} \label{eM1a}
\widehat C_{k\ell}
= \int\hat\sg^2(t)
\exp \biggl\{ - \frac{d(\mathbf{s}_k, \mathbf{s}_\ell)}{\hat\rho(t)}
 \biggr\}\,dt,
\end{equation}
with the estimates $\hat\sg^2(t_j)$ and $\hat\rho(t_j)$ obtained using
some version of empirical variogram, (\ref{egamma-MT}) or (\ref{egamma-HC})
in this study.

Another way to proceed is to replace the $\hat\rho(t_j)$ by their
average
$
\hat\rho= m^{-1}\*\sum_{j=1}^{m}\hat\rho(t_j),
$
where $m$ is the count of the $t_j$ at which the variogram is
estimated successfully. Then, the $C_{k\ell}$ are approximated by
\[
\widehat C_{k\ell}
=  \biggl( \int\hat\sg^2(t)\,dt  \biggr)
\exp \biggl\{ - \frac{d(\mathbf{s}_k, \mathbf{s}_\ell)}{\hat\rho}
\biggr\}.
\]

As explained above, in order to compute the weights $w_j$ in (\ref{ew-mean}),
it is enough to know the matrix $\bC$ only up to a multiplicative
constant. Thus, we may set
%
%e2.7 ###
\begin{equation} \label{eM1b}
\widehat C_{k\ell}
= \exp \biggl\{ - \frac{d(\mathbf{s}_k, \mathbf{s}_\ell)}{\hat\rho}
 \biggr\}.
\end{equation}

Once the matrix $\bC$ has been estimated, we compute the weights
$w_j$, and estimate the mean via (\ref{ehat-mu}).

If (\ref{eM1a}) is used, we refer to this method as M1a; if
(\ref{eM1b}) is used, we call it M1b.

%
%determined using method $M1$, as a function of time. The horizontal
%line is its average value, $\bar{\rho} = 0.474$.} The gaps indicate
%the times $t_j$ where the method failed to converge.
%}
%}%comment

%

Method M1 relies on the estimation of the variograms at every point
$t_j$. Method M2, described below, requires only one optimization, so
it is much faster than M1.

\textit{Method} M2.
We define the \textit{functional} variogram
%
%e2.8 ###
\begin{eqnarray}\label{ef-variog}
2\gamma(\mathbf{s}_k,\mathbf{s}_\ell)
&=& E\| X(\mathbf{s}_k) - X(\mathbf{s}_\ell) \|^2\nn\\
&=&
2E\Vert X(\mathbf{s}_k)-\mu\Vert^2
-2E [\langle X(\mathbf{s}_k)-\mu, X(\mathbf{s}_\ell)-\mu\rangle
 ]
\\
&=& 2E\Vert X(\mathbf{s})-\mu\Vert^2 - 2C_{k\ell}. \nn
\end{eqnarray}
The variogram (\ref{ef-variog}) can be estimated by its empirical
counterparts, like~(\ref{egamma-MT}) or (\ref{egamma-HC}), with
the $|X(\mathbf{s}_k) - X(\mathbf{s}_\ell)|$ replaced by
\[
\|X(\mathbf{s}_k) - X(\mathbf{s}_\ell)\|
=  \biggl\{ \int \bigl( X(\mathbf{s}_k; t) - X(\mathbf{s}_\ell; t) \bigr)^2\,dt  \biggr\}^{1/2}.
\]

Next, we fit a parametric model, for example, we postulate that
%
%e2.9 ###
\begin{equation} \label{ef-exp}
\gamma(\mathbf{s}_k,\mathbf{s}_\ell)
= \sg_f^2  \biggl( 1 - \exp \biggl\{ - \frac{d(\mathbf{s}_k, \mathbf{s}_\ell
)}{\rho_f}  \biggr\} \biggr).
\end{equation}
The subscript $f$ is used to emphasize the \textit{functional} variogram.
Denoting by~$\hat\rho_f$ the resulting NLS estimate, we estimate
the $C_{kl}$ by (\ref{eM1b}) with $\hat\rho$ replaced by $\hat\rho_f$.

\textit{Method} M3.
This method uses a basis expansion
of the functional data, it does not use the weighted
sum (\ref{ehat-mu}).
Suppose $B_j, 1 \le j \le K, $ are elements of a functional
basis with $K$ so large that for each $k$
%
%e2.10 ###
\begin{equation} \label{eexp}
X(\mathbf{s}_k)\approx\sum_{j \le K} \langle B_j, X(\mathbf{s}_k) \rangle B_j
\end{equation}
to a good approximation. By (\ref{e1p}), we
obtain for every $j$
%
%e2.11 ###
\begin{equation} \label{e2p}
\langle B_j, X(\mathbf{s}_k) \rangle = \langle B_j, \mu\rangle+
\langle B_j, \eg(\mathbf{s}_k) \rangle,   \qquad  k =1,2,\ldots, N.
\end{equation}
For every fixed $j$, the $\langle B_j, X(\mathbf{s}_k) \rangle$ form a
stationary and
isotropic scalar spatial field with a constant unknown
mean $\langle B_j, \mu\rangle$.
This mean can be estimated by postulating a covariance structure for
each $\langle B_j, X(\mathbf{s}_k) \rangle$, for example,
\[
\cov (\langle B_j, X(\mathbf{s}_k) \rangle,\langle B_j, X(\mathbf{s}_\ell) \rangle  )
= \sg_j^2 \exp \biggl\{ - \frac{d(\mathbf{s}_k, \mathbf{s}_\ell)}{\rho_j}
 \biggr\}.
\]
The mean $\langle B_j, \mu\rangle$ is estimated by a weighted average
of the $\langle B_j, X(\mathbf{s}_k) \rangle$ (the weights depend on $j$).
Denote the resulting estimate by $\hat\mu_j$.
The mean function $\mu$ is then estimated by
$\hat\mu(t) = \sum_{j \le K} \hat\mu_j B_j(t)$.

%s3 ###
\section{Estimation of
the principal components} \label{sPC}
Assume now that the mean function $\mu$ has been estimated,
and this estimate is subtracted from the data. To simplify
the formulas, in the following we thus assume that $EX(\mathbf{s})=0$.

We consider analogs of methods M2 and M3. Extending Method M1
is possible, but presents a computational challenge because
a parametric spatial model would need to be estimated for every
pair $(t_i, t_j)$. For the ionosonde data studied
in Section \ref{sapp-te}, there are 336 points $t_j$.
Estimation on a single data set would be feasible, but not a simulation
study based on thousands of replications.

In both approaches, which we term CM2 and CM3,
the FPCs are estimated by expansions of the form
%
%e3.1 ###
\begin{equation} \label{evBrep}
v_j(t) = \sum_{\ag=1}^K x_\ag^{(j)} B_\ag(t),\vadjust{\goodbreak}
\end{equation}
where the $ B_\ag$ are elements of an \textit{orthonormal} basis.
We first describe an analog of method M3, which is conceptually and
computationally simpler.

\textit{Method} CM3. The starting point is the expansion
\[
X(\mathbf{s}; t) = \sum_{j=1}^\infty\xi_j(\mathbf{s}) B_j(t),
\]
where, by the orthonormality of the $B_j$,
the $\xi_j(\mathbf{s})$ form an observable field
$\xi_j(\mathbf{s}_k) = \langle B_j, X(\mathbf{s}_k) \rangle$.
Using the orthonormality of the $B_j$ again, we obtain
%
%e3.2 ###
\begin{eqnarray} \label{e3p}
C(B_j)
&=& E  \Biggl[ \Biggl\langle\sum_{\ag=1}^\infty\xi_\ag(\mathbf{s}) B_\ag,
B_j\Biggr\rangle
\sum_{i=1}^\infty\xi_i(\mathbf{s}) B_i \Biggr]\nn\\
&=& E \Biggl [ \xi_j(\mathbf{s}) \sum_{i=1}^\infty\xi_i(\mathbf{s})
B_i \Biggr] \\
&=& \sum_{i=1}^\infty E[\xi_i(\mathbf{s}) \xi_j(\mathbf{s})] B_i. \nn
\end{eqnarray}
Thus, to estimate $C$, we must estimate the means $ E[\xi_i(\mathbf{s})
\xi_j(\mathbf{s})]$.

Fix $i$ and $j$, and define the scalar field $z$ by
$
z(\mathbf{s}) = \xi_i(\mathbf{s}) \xi_j(\mathbf{s}).
$
We can postulate a parametric model for the covariance structure of the
field $z(\cdot)$, and use an empirical variogram to
estimate $\mu_z = E z(\mathbf{s})$ as a weighted average of the $z(\mathbf{s}_k)$.
Denote the resulting estimate by $\hat r_{ij}$.

The empirical version of (\ref{e3p}) is then
%
%e3.3 ###
\begin{equation} \label{e4p}
\widehat C (B_j) = \sum_{i=1}^K \hat r_{ij} B_i.
\end{equation}
Relation (\ref{e4p}) defines the estimator $\widehat C$ which
acts on the span of $B_j$, \mbox{$1\le j \le K$}.

Its eigenfunctions are of the form
$x= \sum_{1 \le\ag\le K} x_\ag B_\ag$.
Observe that
\[
\widehat C (x) = \sum_\ag x_\ag\sum_i \hat r_{i\ag} B_i
= \sum_i  \biggl( \sum_\ag\hat r_{i\ag} x_\ag \biggr) B_i.
\]
On the other hand,
\[
\la x = \sum_i \la x_i B_i.
\]
Since the $B_i$ form an orthonormal basis, we obtain
\[
\sum_\ag\hat r_{i\ag} x_\ag= \la x_i.
\]
Setting
\[
\bx= [x_1, x_2, \ldots, x_K]^T,    \qquad
\widehat\bR= [ \hat r_{ij}, 1 \le i, j \le K],\vadjust{\goodbreak}
\]
we can write the above as a matrix equation
%
%e3.4 ###
\begin{equation} \label{e5p}
\widehat\bR\bx= \la\bx.
\end{equation}
Denote the solutions to (\ref{e5p}) by
%
%e3.5 ###
\begin{equation} \label{e6p}
\hat{\mathbf x}^{(j)}
= \bigl[ \hat x^{(j)}_1, \hat x^{(j)}_2, \ldots, \hat x^{(j)}_k\bigr]^T,   \qquad
\hla_j, \qquad
   1 \le j \le K.
\end{equation}
The $\hat{\mathbf x}^{(j)}$ satisfy
$ \sum_{\ag=1}^K \hat x^{(j)}_\ag\hat x^{(i)}_\ag=\dg_{ij}$.
Therefore, the $\hv_j$ defined by
%
%e3.6 ###
\begin{equation} \label{e7p}
\hv_j = \sum_{\ag=1}^K \hat x^{(j)}_\ag B_\ag
\end{equation}
are also orthonormal (because the $B_j$ are orthonormal).
The $\hv_j$ given by~(\ref{e7p}) are the estimators of the FPCs, and the
$\hla_j$ in (\ref{e6p}) of the corresponding eigenvalues.

As in method M3, the value of $K$ can be taken to the number of basis
functions used to create the functional objects in \texttt{R}, so it can be
a relatively large number, for example, $K=49$. Even though the range
of $j$ in~(\ref{e6p}) and~(\ref{e7p}) runs up to $K$, only the first few
estimated FPCs $\hv_j$ would be used in further work.

\textit{Method} CM2. Recall that under the assumption of zero mean
function, the covariance operator is defined by
$
C(x) = E[\langle X(\mathbf{s}), x \rangle X(\mathbf{s})].
$
It can be estimated by the simple average
%
%e3.7 ###
\begin{equation} \label{eemp-C}
\frac{1}{N} \sum_{n=1}^N \langle X(\mathbf{s}_n), \cdot\rangle X(\mathbf{s}_n)
= \frac{1}{N} \sum_{n=1}^N C_k,
\end{equation}
where $C_k$ is the operator defined by
\[
C_k(x) = \langle X(\mathbf{s}_k), x \rangle X(\mathbf{s}_k).
\]
As for the mean, more precise estimates can be obtained
by using the weighted average
%
%e3.8 ###
\begin{equation} \label{ehat-C}
\widehat C = \sum_{k=1}^N w_k C_k.
\end{equation}

Before discussing the estimation of the weights $w_k$, we comment
that the FPCs $v_j$ and their eigenvalues $\la_j$ can be estimated
using (\ref{ehat-C}) and the representation (\ref{evBrep}).
As in method CM3, set $x= \sum_{1 \le\ag\le K} x_\ag B_\ag$,
and observe that
\[
\widehat C (x) = \sum_{j=1}^K
 \Biggl( \sum_{\ag=1}^K s_{j\ag} x_\ag \Biggr) B_j,
\]
where
\[
s_{j\ag} = \sum_{k=1}^N w_k \langle X_k, B_j \rangle
\langle X_k, B_\ag\rangle.\vadjust{\goodbreak}
\]
Thus, analogously to (\ref{e5p}), we obtain a matrix equation
$\bS\bx= \la\bx$, from which the estimates of the $v_j, \la_j$
can be found as in (\ref{e6p}) and (\ref{e7p}).

We now return to the estimation of the weights $w_k$ in (\ref{ehat-C}).
One way to define the optimal weights
is to require that they minimize the expected
Hilbert--Schmidt norm of $\widehat C - C$. Recall that the
Hilbert--Schmidt norm of an operator $K$ is defined by
\[
\| K\|_{\cS}^2 = \sum_{i=1}^\infty\| K(e_i)\|^2
= \sum_{i=1}^\infty\int|K(e_i)(t)|^2\,dt,
\]
where $ \{ e_i, i \ge1  \}$ is any orthonormal basis in $L^2$.
Since $\|  \cdot \|_{\cS}$ is
a norm in the the Hilbert space $\cS$ of the Hilbert--Schmidt operators
with the inner product
\[
\langle K_1, K_2 \rangle_{\cS} = \sum_{i=1}^\infty\langle K_1(e_i),
K_2(e_i) \rangle,
\]
we can repeat all algebraic manipulations needed to obtain the weight
$w_i$ in (\ref{ehat-mu}). The optimal weights in (\ref{ehat-C})
thus satisfy
%
%e3.9 ###
\begin{equation} \label{ew-C}
\sum_{n=1}^N w_n =1,     \qquad
\sum_{k=1}^N w_k \kg_{kn} -r = 0,    \qquad  n=1,2, \ldots, N,
\end{equation}
where
\[
\kg_{k\ell} = E [ \langle C_k - C, C_\ell- C \rangle_{\cS} ].
\]
Finding the weights thus reduces to estimating the expected
inner products~$\kg_{k\ell}$.

Since method M2 of Section \ref{smean} relies only on
estimating inner product in the Hilbert space $L^2$, it can be
extended to the Hilbert space $\cS$. First observe
that, analogously to (\ref{ef-variog}),
\[
E\| C_k - C_\ell\|_{\cS}^2 = 2E\| C_k -C\|_{\cS}^2 - 2\kg_{k\ell}.
\]
We can estimate the variogram
\[
\ga_C(d) = E\|\langle X(\mathbf{s}), \cdot  \rangle X(\mathbf{s})
- \langle X(\mathbf{s}+\mathbf{d} ), \cdot  \rangle X(\mathbf{s}+\mathbf{d})
\|_{\cS}^2,
  \qquad d = \|\mathbf{d}\|
\]
by fitting a parametric model. In formulas
(\ref{egamma-MT}) and (\ref{egamma-HC}), the squared distances
$(X(\mathbf{s}_k) - X(\mathbf{s}_\ell))^2$ must be replaced by the squared norms
$\| C_k - C_\ell\|_{\cS}^2$.
These norms are equal to
\[
\| C_k - C_\ell\|_{\cS}^2 = \sum_{i=1}^\infty\int
 \bigl( f_{ik} X_k(t) - f_{i\ell} X_\ell(t) \bigr)^2\,dt,
\]
where
\[
f_{ik} = \int X_k(t) e_i(t)\,dt.
\]
The inner products $f_{ik}$ can be computed using
the \texttt{R} package \texttt{fda}.\vadjust{\goodbreak}

%s4 ###
\section{Finite sample performance
of the estimators}
\label{sfs}
In this section we report the results of a simulation study
designed to compare the performance of the methods proposed in
Sections \ref{smean} and \ref{sPC}
in a realistic setting motivated by the ionosonde data.
It is difficult to design
an exhaustive simulation study due to the number of possible combinations
of the point distributions, dependence structures, shapes of mean
functions and the FPCs and ways of implementing the methods
(choice of spatial models, variogram estimation etc.). We do, however,
think that our study provides useful information and guidance
for practical application of the proposed methodology.

\textit{Data generating processes.}
We generate functional data at location $\mathbf{s}_k$ as
%
%e4.1 ###
\begin{equation} \label{esim} X(\mathbf{s}_k; t) = \mu(t) + \sum_{i=1}^p
\xi_{i}(\mathbf{s}_k) e_{i}(t),
\end{equation}
where the $e_i$ are orthonormal functions; cf. model (\ref{emodel}).

To evaluate the estimators of the mean, we use $p=2$ and
%
%e4.2 ###
\begin{equation} \label{ee12}
e_{1}(t) = \sqrt{2}\sin(2\pi t \cdot6),
 \qquad
e_{2}(t) = \sqrt{2}\sin(2\pi t/2).
\end{equation}
We use two mean functions
%
%e4.3 ###
\begin{equation} \label{emu-io}
\mu(t) = a \sqrt{2}\sin(2\pi t \cdot3),    \qquad a= 2,
\end{equation}
and
%
%e4.4 ###
\begin{equation} \label{emu-sq}
\mu(t) = a \sqrt{t}\sin(2\pi t \cdot3),    \qquad  a= 1.
\end{equation}

The mean function (\ref{emu-io}) resembles the mean shape for the
ionosonde data. It is, however, a member of the Fourier
basis, and can be isolated using only one basis function, what
could possibly artificially enhance the performance of method M3. We
therefore also consider the mean function~(\ref{emu-sq}). Combining
the mean function~(\ref{emu-io}) and the FPCs (\ref{ee12}), we obtain
functions which very closely resemble the shapes of the ionosonde curves.
In the above formulas, time is rescaled so that $t \in[0,1]$.

To evaluate the estimators of the FPCs, we use $p=3$ and
%
%e4.5 ###
\begin{equation} \label{esim-FPC}
X(\mathbf{s}_k;t)=\xi_1(\mathbf{s}_k)\frac{e_1(t)+e_2(t)}{\sqrt{2}}+\xi
_2(\mathbf{s}_k)e_3(t),
\end{equation}
where $e_1(t) =\sqrt{2}\sin(2\pi t\cdot7)$,
$e_2(t) =\sqrt{2}\sin(2\pi t\cdot2)$,
$e_3(t) =\sqrt{2}\sin(3\pi t\cdot3)$.
Direct verification, which uses the independence of the fields
$\xi_1$ and $\xi_2$, shows that the FPCs are
$v_1 = 2^{-1/2}(e_1+ e_2)$ and $v_2 = 2_3$ (for the parameters
of the~$\xi_i$ specified below).

To complete the description of the data generating processes,
we must
specify the dependence structure of the scalar spatial fields $\xi_1$
and $\xi_2$. A~common assumption for the
Karhunen--Lo{\'e}ve expansions used in statistical inference is that
the score processes $\xi_i$ are independent, and this is what we
assume. We\vadjust{\goodbreak} use the exponential and Gaussian models (\ref{eE-G})
with chordal distances (\ref{edistance})
between the locations described below.
To make simulated data look
similar to the real foF2 data, we chose $\sg_1 = 1$,
$\rho_1=\pi/6$ for $\xi_1(\mathbf{s})$ field and $\sg_2 = 0.1$, $\rho
_2 = \pi/4$
for $\xi_2(\mathbf{s})$ field.

The locations $\mathbf{s}_k$ are selected to match the locations of the real
ionosonde stations. For the sample size $218$ we use all available
locations, as shown in Figure~\ref{floc}. Size 32 corresponds to the
ionosondes with the longest record history. We also consider a sample
of size 100; the 100 stations were selected randomly out of the 218
stations.

\textit{Details of implementation.} All methods require the specification
of a parametric spatial model for the variogram. Even though for some
methods the variograms are defined for $L^2$- or $\cS$-valued
objects, only a \textit{scalar} model is required. In this simulation
study we use the exponential and Gaussian models.\looseness=-1

Methods M3, CM2 and CM3 require the specification of a basis $\{ B_j\}$
and the number $K$ of the basis functions. We use the Fourier
basis and $K=1+4[\sqrt{\#\{ t_j\}}]$, where $\#\{ t_j\}$ is the count of
the points at which the curves are observed. For our real and simulated data
$K=1 + 4[\sqrt{336}]= 73$, a number that falls between the recommended
values of 49 and 99 for the number of basis functions.
Specifically, the basis functions $B_j$ are
%
%e4.6 ###
\begin{equation} \label{ef-basis}
\bigl\{1, \sqrt{2}\sin(2\pi t \cdot i), \sqrt{2}\cos(2\pi t \cdot i);
   i = 1,2,\ldots, 36 \bigr\}.
\end{equation}

All methods require the estimation of a parametric model on an
empirical variogram.
There are several versions of the empirical variogram for scalar fields.
The classical estimator proposed by Matheron is given by
%
%e4.7 ###
\begin{equation}
\label{egamma-MT}
\hat{\gamma}(d)
= \frac{1}{|N(d)|}\sum_{N(d)}\bigl(X(\mathbf{s}_k)-X(\mathbf{s}_l)\bigr)^2,
\end{equation}
where $N(d) =  \{ (\mathbf{s}_i, \mathbf{s}_j)\dvtx  d_{\mathbf{s}_i,\mathbf{s}_j}
= d; i,j = 1,\ldots,
N \}$ and $|N(d)|$ is the number of distinct pairs in $N(d)$.
A robust estimator proposed by Cressie and Hawkins is defined as
%
%e4.8 ###
\begin{equation}
\label{egamma-HC}
\hat{\gamma}(d)
=  \biggl(\frac{1}{|N(d)|}\sum_{N(d)}
|X(\mathbf{s}_k)-X(\mathbf{s}_l)|^{1/2} \biggr)^4 \bigg/ \biggl(0.457+\frac
{0.494}{|N(d)|} \biggr).
\end{equation}
For details, we refer to Section 4.4 of Schabenberger and Gotway
(\citeyear{schabenberger-gotway-2005}),
where other ways of variogram estimation are also discussed.
In our study we
use only estimators (\ref{egamma-MT}) and (\ref{egamma-HC}),
and refer to them, respectively, as MT and CH.

\textit{Results of the simulation study.}
For comparison of different methods we introduce the quantity $L$ which is
the average of the integrated absolute differences between real and
estimated mean functions or FPCs. For the mean function, $L$ is defined
by
%
%e4.9 ###
\begin{equation} \label{eL}
L = \frac{1}{R}\sum_{r=1}^{R}\int|\hat{\mu}_r(t)-\mu(t)|\,dt,\vadjust{\goodbreak}
\end{equation}
where $R$ is the number of replications; we use $R=10^3$.
For the FPCs the definition is fully analogous. We also compute
the standard deviation for $L$, based on the normal approximation
for $R$ independent runs.

%f4 ###
\begin{figure}

\includegraphics{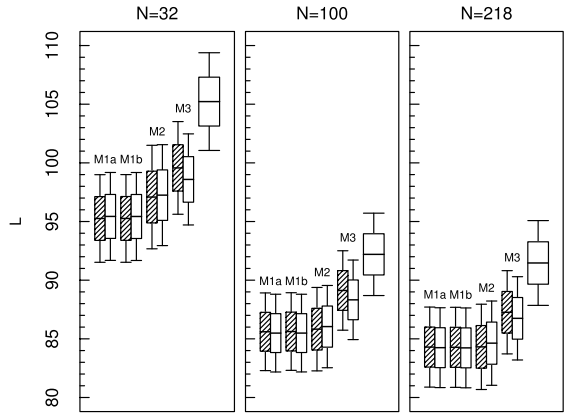}

\caption{Errors in the estimation of the mean function for sample
sizes: $32$,
$100$, $218$. The dashed boxes are estimates using the CH variogram,
empty are for the MT variogram. The rightmost box for each $N$ corresponds
to the simple average. The bold line inside each box plot represents
the average value of $L$ (\protect\ref{eL}). The upper and lower sides of rectangles
show one standard deviation, and horizontal lines show two standard deviations.
The rightmost boxes correspond to the standard method.}
\label{fcmean}
\end{figure}

The results of the simulations for the mean function (\ref{emu-sq})
are shown in Figure~\ref{fcmean}. The data generating processes have
exponential covariance functions. If the $\xi_i$ in (\ref{esim})
have Gaussian covariances, the results are not visually distinguishable.
The errors values for mean (\ref{emu-io}) are slightly different,
but the relative position of the box plots practically does not change.
All methods M1, M2 and M3 are significantly better than the sample
average. Method M2 strikes the best balance between the computational
cost and the precision of estimation. Note that methods M1 and M2 were
designed to minimizes the expected $L^2$
distance, and all three methods are compared using the $L^1$ distance,
so this
comparison does not {a priori} favor them.
In the context of forecasting, using different loss functions to evaluate
the forecasts than to design them can lead to spurious conclusions;
see Gneiting (\citeyear{gneiting-2011}).
In our context, if the $L^2$ distance is
used to compare the methods, the ranking and conclusions are the same.

%f5 ###
\begin{figure}

\includegraphics{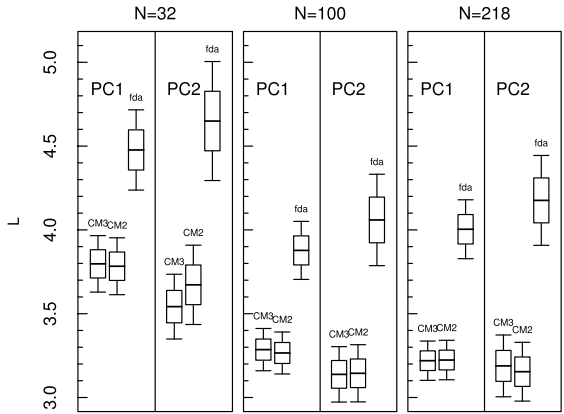}

\caption{Errors in the estimation of the FPCs for sample sizes: $32$,
$100$, $218$. The bold line inside each box plot represents the average
value of $L$. The upper and lower sides of rectangles
show one standard deviation, and horizontal lines show
two standard deviations.
The rightmost boxes correspond to the standard method.}
\label{fpca}
\end{figure}

Errors in the estimation of the FPCs in model (\ref{esim-FPC}) are
shown in Figure~\ref{fpca}. The displayed errors are those for the
$\xi_i$ with exponential covariances and the CH variogram. The results
for Gaussian covariances and the MT variogram are practically the
same. The performance of methods CM2 and CM3 is comparable, and they
are both much better than using the eigenfunctions of the empirical
covariance operator (\ref{eemp-C}), which is the standard method
implemented in the \textsf{fda} package. The computational complexity
of methods CM2 and CM3 is the same.

\textit{Conclusions.} For simulated data generated to resemble the
ionosonde data, all methods introduced in Sections \ref{smean} and
\ref{sPC} have integrated absolute deviations (away from a true
curve) statistically significantly smaller than the standard methods
designed for i.i.d. curves. Methods M2 and CM2, based on weighted
averages estimated using functional variograms, offer a~computationally efficient and unified approach to the estimation of
the mean function and of the FPCs in this spatial setting.

%s5 ###
\section{Testing for correlation of
two spatial fields} \label{ste}
Motivated by the problem of testing for correlation between
foF2 and magnetic curves, described in detail in Section \ref{sapp-te},
we now propose a relevant statistical significance test.

There are $N$ spatial locations: $\mathbf{s}_1, \mathbf{s}_2, \ldots, \mathbf{s}_N$.
At location $\mathbf{s}_k$, we have two curves:
\[
X_k = X(\mathbf{s}_k) = X(\mathbf{s}_k; t), \qquad  t \in[0,1],
\]
and
\[
Y_k = Y(\mathbf{s}_k) = Y(\mathbf{s}_k; t), \qquad  t \in[0,1].
\]

We want to test the null hypothesis that the collections of curves
$\{X_k,  1\le k \le N\}$ and
$\{Y_k,  1\le k \le N\}$ are uncorrelated in a sense defined below.
The null distribution is derived assuming a stronger condition
that these two families are independent.
Sz\'ekely, Rizzo and Bakirov (\citeyear{szekely-rizzo-bakirov-2007})
and Sz\'ekely and Rizzo (\citeyear{szekely-rizzo-2009}) introduced
measures of dependence based on the distance of
characteristic\vadjust{\goodbreak} functions which allow to test independence (rather than
just lack of correlation) of
random variables $X$ and $Y$ given a sample of i.i.d. observations
$(X_k, Y_k)$. The extension of their theory to the case of spatially
dependent observations $(X_k, Y_k)$ is not obvious, so we consider
only a test for linear dependence.

The idea of the test is as follows. To lighten the notation, assume that
\[
EX_k(t)=0    \quad \mbox{and}  \quad     EY_n(t)=0.
\]
The mean functions will be estimated and subtracted using
one of the methods of Section~\ref{smean}.
We approximate the curves $X_n$ and $Y_n$ by
the expansions
\[
X_n(t) \approx\sum_{i=1}^p \langle X_n, v_i \rangle v_i(t),    \qquad
Y_n(t) \approx\sum_{j=1}^q \langle Y_n, u_j \rangle u_j(t),
\]
where the $v_i$ and the $u_j$ are the corresponding FPCs.
At this point,
the functions $v_i,  1 \le i \le p$ and $u_j,  1 \le j \le q$
are deterministic, so
the independence of the curves $X_n$ of the curves $Y_n$
implies the independence of the vectors
\[
[ \langle X_n, v_1 \rangle, \langle X_n, v_2 \rangle, \ldots,
\langle X_n, v_p \rangle]^T,
   \qquad  1 \le n \le N
\]
and
\[
[ \langle Y_n, u_1 \rangle, \langle Y_n, u_2 \rangle, \ldots,
\langle Y_n, u_q \rangle]^T,
   \qquad  1 \le n \le N.
\]
Then, under $H_0$, the expected value of the
sample covariances
%
%e5.1 ###
\begin{equation}
\label{eAn}
A_N(i,j) = \frac{1}{N}\sum_{n=1}^{N} \langle X_n,
{v}_i \rangle\langle Y_n, {u}_j \rangle
\end{equation}
is zero. If their estimated versions are large as a group,
that is, if some of the estimated $A_N(i,j)$ are too large, we reject
the null
hypothesis.

To construct a test statistic, we introduce the quantities
\[
V_{k\ell}(i,\ip) = E[\langle v_i, X_k \rangle\langle v_\ip, X_\ell
\rangle],
  \qquad
U_{k\ell}(j,\jp) = E[\langle u_j, Y_k \rangle\langle u_\jp, Y_\ell
\rangle].
\]
Note that $V_{k\ell}(i,\ip)=0$ and $U_{k\ell}(j,\jp)=0$,
if the observations in each sample are independent (and have
mean zero). Thus, the $V_{k\ell}(i,\ip)$ and the $U_{k\ell}(j,\jp)$
are specific to dependent data, they do not occur in the currently
available testing procedures developed for independent curves.
Setting
$
X_{ik} = \langle v_i, X_k \rangle,   Y_{jk}= \langle u_j, Y_k \rangle,
$
observe that if the $X_{ik}$ are uncorrelated with the
$Y_{jk}$, then
\begin{eqnarray*}
E\bigl[ \sqrt{N} A_N(i,j)\sqrt{N} A_N(\ip,\jp)\bigr]
&=& \frac{1}{N}
E  \Biggl[ \sum_{k=1}^N X_{ik} Y_{jk} \sum_{\ell=1}^N X_{\ip\ell}
Y_{\jp\ell}  \Biggr]
\\
&=& \frac{1}{N} \sum_{k=1}^N\sum_{\ell=1}^N
E [X_{ik} X_{\ip\ell}] E[ Y_{jk} Y_{\jp\ell}]
\\
&=& \frac{1}{N} \sum_{k=1}^N\sum_{\ell=1}^N
V_{k\ell}(i,\ip)U_{k\ell}(j,\jp).
\end{eqnarray*}
The normalized covariance tensor of the $\sqrt{N} A_N(i,j)$ thus
has entries
%
%e5.2 ###
\begin{equation} \label{ecov-ten}
\sg_N(i,j; \ip, \jp) =
\frac{1}{N} \sum_{k,\ell=1}^N V_{k\ell}(i,\ip)U_{k\ell}(j, \jp).
\end{equation}
The idea of the test is to approximate the distribution of the matrix
\[
\bA_N = [ A_N(i,j),  1 \le i \le p,  1 \le j \le q ]
\]
via
$
\sqrt{N} \bA_N \approx\bZ,
$
where $\bZ$ is a $p\times q$ Gaussian matrix whose elements have
covariances
$
E[Z(i,j) Z(\ip, \jp)] = \sg_N(i,j; \ip, \jp).
$

We now explain how to implement this idea.
Denote by $\hat\la_i, \hat\ga_j$ and $\hv_i, \hu_j$
the eigenvalues and the eigenfunctions estimated either by method CM2
or CM3. The covariances $A_N(i,j)$ are then estimated by
\[
\hat A_N(i,j) = \frac{1}{N}\sum_{n=1}^{N} \langle X_n,
\hv_i \rangle\langle Y_n, \hu_j \rangle.
\]

If the observations within each sample are independent,
an appropriate test statistic is
\[
N \sum_{i=1}^p \sum_{j=1}^q {\hat\la_i}^{-1}{\hat\ga_j}^{-1}
\hat A_N^2(i,j).
\]
Since $\la_i = E [\langle v_i, X \rangle^2]$, this is essentially the sum
of all correlations, and it tends to a chi-squared distribution with
$pq$ degrees of freedom, as shown in Kokoszka et al. (\citeyear
{kokoszka-maslova-s-z-2008}).
This is, however, not that case for dependent data. To explain,
set
\[
\mathbf{a}_N = \operatorname{vec} (\bA_N),
\]
that is, $\mathbf{a}_N$ is a column vector of length $pq$ consisting of
the columns of $\bA_N$ stacked on top of each other,
starting with the first column. Then
$\sqrt{N} \mathbf{a}_N$ is approximated by a Gaussian vector $\mathbf{z}$
with covariance
matrix $\bSig$ constructed from the entries (\ref{ecov-ten}).
It follows that
%
%e5.3 ###
\begin{equation} \label{eSN}
\hat S_N = N {\hat\mathbf{a}_N}^T {\hat\bSig}^{-1}{\hat\mathbf{a}_N}\approx\chi^2_{pq},
\end{equation}
where $\hat\mathbf{a}_N = \operatorname{vec}(\hat\bA_N)$.
The entries of the matrix $\hat\bSig$ are
%
%e5.4 ###
\begin{equation} \label{ehat-sg}
\hat\sg_N(i,j; \ip, \jp) = \frac{1}{N} \sum_{k,\ell=1}^N
{\hat V}_{k\ell}(i,\ip) {\hat U}_{k\ell}(j, \jp),
\end{equation}
where ${\hat V}_{k\ell}(i,\ip)$ and ${\hat U}_{k\ell}(j, \jp)$ are
estimators of ${ V}_{k\ell}(i,\ip)$ and ${U}_{k\ell}(j, \jp)$,
respectively.
The test rejects $H_0$ if $\hat S_N > \chi^2_{pq}(1-\ag)$, where
$\chi^2_{pq}(1-\ag)$ is the 100$(1-\ag)$th percentile of the
chi-squared distribution with $pq$ degrees of freedom.
One can use Monte Carlo versions of the above test, for example,
the test is based on
the approximation
%
%e5.5 ###
\begin{equation} \label{eTN}
\hat T_N := N {\hat\mathbf{a}_N}^T {\hat\mathbf{a}_N} \approx\bw^T {\hat
\bSig} \bw,
\end{equation}
where the components of $\bw$ are i.i.d. standard normal.

The test procedure can be summarized as follows:
\begin{longlist}[(3)]
\item[(1)] Subtract the mean functions, estimated by
one of the methods of Section \ref{smean}, from both samples.
\item[(2)] Estimate the FPCs by method CM2 or CM3.
\item[(3)] Using a model for the covariance tensor (\ref{ehat-sg}) (see
Section~\ref{sestSig}), compute the test statistic $\hat S_N$.
(This tensor is not needed to compute $\hat T_N$, but it is needed
to find its Monte Carlo distribution.)
\item[(4)] Find the $P$-value using either a Monte Carlo distribution
or the $\chi^2$ approximation.
\end{longlist}

We now turn to the important issue of modeling and estimation
of the matrix $\bSig$.

%s6 ###
\section{Modeling and estimation
of the covariance tensor} \label{sestSig}

The estimation of the $V_{k\ell}(i,\ip)$ involves
only the $X_n$, and the estimation of the $U_{k\ell}(j, \jp)$ only
the $Y_n$,
so we describe only the procedure for the $V_{k\ell}(i,\ip)$.
We assume that the mean has been estimated and subtracted,
so that we can define
%
%e6.1 ###
\begin{equation} \label{eCh}
C_h(x) = E[ \langle X(\mathbf{s}), x \rangle X(\mathbf{s}+ \bh) ],    \qquad  h =
\|\bh\|.
\end{equation}
The estimation of the $V_{k\ell}(i, \ip)$ relies on the identity
\[
V_{k\ell}(i, \ip) = \langle C_h(v_i), v_{\ip} \rangle,     \qquad
h = d(\mathbf{s}_k, \mathbf{s}_\ell).
\]

To propose a practical approach to the estimation of $\bSig$,
we consider an extension of the multivariate
intrinsic model; see, for example, Chapter 22 of Wackernagel (\citeyear
{wackernagel-2003}).
A most direct extension is to assume that
%
%e6.2 ###
\begin{equation} \label{eintrinsic}
C_h = C r(h),
\end{equation}
where $C$ is a covariance operator, that is, a symmetric positive
definite operator with summable eigenvalues, and $r(h)$
is a correlation function of a scalar random field.
Since $r(0) = 1$, we have $C=C_0$, so $C$ in (\ref{eintrinsic})
must be the covariance operator of each $X(\mathbf{s})$.
If we assume the intrinsic model~(\ref{eintrinsic}), then
%
%e6.3 ###
\begin{equation} \label{eint}
V_{k\ell}(i,j) = \langle r(h) C(v_i), v_{j} \rangle
= \la_i \dg_{ij} r(d(\mathbf{s}_k, \mathbf{s}_l)).
\end{equation}
To allow more modeling flexibility, we postulate that
%
%e6.4 ###
\begin{equation} \label{eint-mod}
V_{k\ell}(i,j) = \la_i \dg_{ij} r_i(d(\mathbf{s}_k, \mathbf{s}_l)).
\end{equation}
Under (\ref{eint}) [equivalently, under (\ref{eintrinsic})],
each scalar field $\langle X(\mathbf{s}), v_i \rangle$
has the same correlation function, only their variances are different.
Under (\ref{eint-mod}), the fields $\langle X(\mathbf{s}), v_i \rangle$
can have
different correlation functions. As will be seen below, model
(\ref{eint-mod}) also leads to a valid covariance matrix.

The correlations $r_i(d(\mathbf{s}_k, \mathbf{s}_l))$ and the variances
$\la_i$ can be estimated
using a parametric model for the scalar field
$\xi_i(\mathbf{s}) = \langle X(\mathbf{s}), v_i \rangle$. The resulting
estimates $\hat r_i(d(\mathbf{s}_k, \mathbf{s}_l))$ and
$\hla_i$ lead to the estimates
$\hat V_{k\ell}(i,j)$ via (\ref{eint-mod}). Analogous estimates
of the functional field $Y$ are $\hat\ga_j(d(\mathbf{s}_k, \mathbf{s}_l)),
\hat\tau_j$ and
$\hat U_{k\ell}(i,j)$.

For ease of reference,
we note that under model (\ref{eint-mod}) and $H_0$,
the covariance tensor,
\[
 \Biggl[\frac{1}{N}\sum_{k=1}^{N}\sum_{\ell=1}^{N}
\hat{V}_{k\ell}(i,i')\hat{U}_{k\ell}(j,j'),
1\leq i,i' \leq p,   1\leq j,j' \leq q \Biggr]  ,
\]
has the following matrix representation:
%
%e6.5 ###
\begin{equation}
\label{eSig-S}
\hat{\bSig} = \operatorname{diag} \Biggl( \sum_{k=1}^{N}\sum_{\ell
=1}^{N}\hat{\bSig}_{\xi_1}(k,\ell)\hat{\bSig}_{\eta_1}(k,\ell
),\ldots,
\sum_{k=1}^{N}\sum_{\ell=1}^{N}\hat{\bSig}_{\xi_p}(k,\ell)\hat
{\bSig}_{\eta_q}(k,\ell)
 \Biggr) ,\hspace*{-30pt}
\end{equation}
where
\[
\hat{\bSig}_{\xi_i}(k,\ell) = \frac{1}{\sqrt{N}}\hat{\lambda}_i
\hat{r}_i (d(\mathbf{s}_k,\mathbf{s}_{\ell}))
\]
and
\[
\hat{\bSig}_{\eta_j}(k,\ell) = \frac{1}{\sqrt{N}}\hat{\gamma}_j
\hat{\tau}_j (d(\mathbf{s}_k,\mathbf{s}_{\ell})).
\]
This form is used to construct the Monte Carlo tests discussed
in Section \ref{ssp-cor}.

The matrices $\bSig$ and $\widehat\bSig$ are
positive definite; see Horv\'ath and Kokoszka (\citeyear{HKbook}) for
the verification.

%s7 ###
\section{Size and power of the correlation test} \label{ssp-cor}
As in Section \ref{sfs}, our objective is
to evaluate the finite sample performance of the test introduced in
Section~\ref{ste} in a realistic setting geared toward the application
presented in Section~\ref{sapp-te}.

%t1 ###
\begin{table}
\tabcolsep=0pt
\caption{Models and estimated covariance parameters for the transformed
foF2 curves and the magnetic curves}
\label{tbres}
\begin{tabular*}{\textwidth}{@{\extracolsep{\fill}}lcccc@{}}
\hline
 &  & \multicolumn{3}{c@{}}{\textbf{Parameters}}\\[-5pt]
\textbf{Spatial}  &   & \multicolumn{3}{c@{}}{\hrulefill}\\
\textbf{field} & \textbf{Model}& $\boldsymbol{c_0}$ & $\boldsymbol{\sigma^2}$ & $\boldsymbol{\rho}$\\
\hline
$\bfeta$ & Gaussian & -- & $5.99 \pm0.48$ & $0.32\pm0.04$\\
$\bxi_1$ & Gaussian & -- & $20.05 \pm2.20$ & $0.12\pm0.03$\\
$\bxi_2$ & -- & -- & $3.30 \pm0.43$ & -- \\
$\bxi_3$ & Exponential & -- & $2.63 \pm0.52$ & $0.16\pm0.07$\\
$\bxi_4$ & Gaussian & -- & $2.66 \pm0.39$ & $0.18\pm0.05$\\
$\bxi_5$ & -- & -- & $2.74 \pm0.32$ & -- \\
$\bxi_6$ & Gaussian & $0.16\pm0.02$ & $0.85 \pm0.24$ & $0.17\pm
0.06$\\
$\bxi_7$ & -- & -- & $1.22 \pm0.18$ & -- \\
\hline
\end{tabular*}
\end{table}

\textit{Data generating processes.}
We generate samples of zero mean Gaussian processes
%
%e7.1 ###
\begin{equation}
\label{exy-sim}
X(\mathbf{s}; t) = \sum_{i = 1}^{p} \xi_i(\mathbf{s})v_{i}(t);
   \qquad
Y(\mathbf{s}; t) = \sum_{j=1}^{q}\eta_j(\mathbf{s})u_j(t).
\end{equation}
The process $X$ is designed to resemble in distribution appropriately
transformed and centered foF2 curves; the process $Y$ the centered
magnetic curves.
Following the derivation presented in Section~\ref{sapp-te}, we use
$p=7$ and $q=1$. The curves $v_i$ and $u_1$ are the estimated FPCs
of the real data. The scalar Gaussian spatial fields $\xi_i$ and $\eta_1$
follow parametric models estimated for real data; details of the models
are presented in Table~\ref{tbres}. The $\xi_i$ are independent.
Under~$H_0$, the $\xi_i$ are independent of $\eta_1$. The dependence under~$H_A$ can be generated in many ways. We considered the following
scenarios:~$\xi_1$ and~$\eta_1$ are dependent,~$\xi_i$ and~$\eta_1$ are independent
for $i\neq1$, then~$\xi_2$ and~$\eta_1$ are dependent,~$\xi_i$ and~$\eta_1$ are independent
for $i\neq2$, etc.
To produce two dependent spatial fields~$\xi_i$ and~$\eta$,
we generated $N$ i.i.d. pairs $\bx_i=[x_{1i}, x_{2i}]^T$,
$1\leq i \leq N$, where
\[
\bx_i\sim N \left(\mathbf{0},
\pmatrix{
1& \rho\cr
\rho& 1
}
 \right) .
\]
Then we merged all $x_{1i}$ into vector $\by_1 =
[x_{11},\ldots, x_{1N}]^T $ and all $x_{2i}$ into vector $\by_2 =
[x_{21},\ldots, x_{2N}]^T $. Performing the Cholesky rotation, we obtain
correlated spatial vectors:
\[
\bxi_i = \bV\by_1,    \qquad  (\bSig_{\xi_i} = \bV\bV^T),
  \qquad
\bfeta= \bU\by_2,    \qquad  (\bSig_{\eta} = \bU\bU^T).
\]

We used sample sizes $N=32$ and $N=100$ corresponding to the
locations determined as in Section~\ref{sfs}.

\textit{Testing procedures.} We studied the finite sample behavior of
three methods, which we call S, SM and T. Method S rejects $H_0$ if
the statistic $\hat S_N$ (\ref{eSN}) exceeds a chi-square critical
value. Method SM uses a Monte Carlo distribution of the statistic
$\hat S_N$: after estimating all parameters from the data and assuming
the Gaussian distribution of the fields $\xi_i$ and $\eta_1$, we can
replicate the values of the statistic $\hat S_N$ under $H_0$ using the
covariance matrix (\ref{eSig-S}). Method T uses the statistics $\hat
T_N$ (\ref{eTN}), and approximates its distribution by the Monte Carlo
distribution of $\bw^T \hat\bSig\bw$, as explained in
Section~\ref{ste}. For determining the critical values
in methods SM and T, we
used $10^7$ Monte Carlo replications. The empirical size and power
are based on $10^5$ independent runs.\vadjust{\goodbreak}

%f6 ###
\begin{figure}

\includegraphics{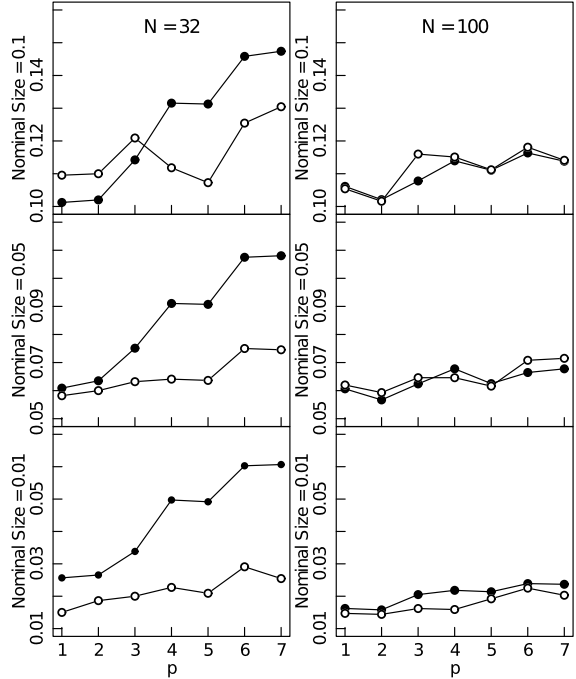}

\caption{Size of the correlation test as a function of $p$.
Solid disks represent method S (based on $\chi^2$ distribution).
Circles represent method SM (based on the Monte Carlo distribution).}
\label{fsize}
\vspace*{6pt}
\end{figure}

\begin{figure}

\includegraphics{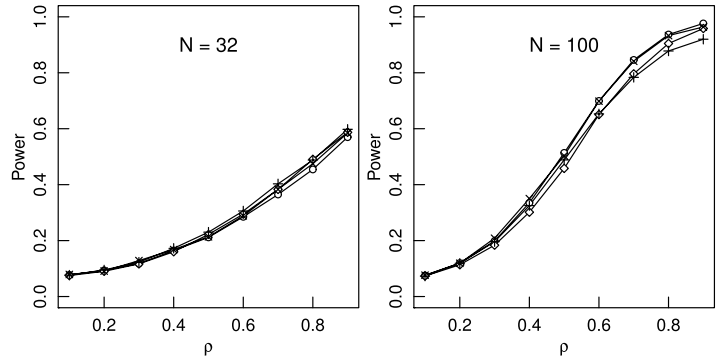}

\caption{Power of the correlation test SM as a function of the population
correlation $\rho$. Each line represents one of the four possible
correlated spatial fields: $\bxi_1 - \bfeta$,
$\bxi_2 - \bfeta$, $\bxi_3 - \bfeta$, $\bxi_4 - \bfeta$.
The test was performed using $p=4$, which explains about $85\%$
of variance of the foF2 curves. Since all curves in the graphs
are practically the same, we do not specify
which curve represents a particular dependent pair $\bxi_i - \bfeta$.}\label{fPower}
\end{figure}

\textit{Conclusions.} As Figure~\ref{fsize} shows, the empirical size is
higher than the nominal size, and it tends to increase with the number
$p$ of principal components used to construct the test, especially
for $N=32$. The usual recommendation is to use $p$, which explains
about 85\% of the variance. For the foF2 data with $N=32$, this
corresponds to $p=4$.
Applied to real data in Section~\ref{sapp-te},
all tests (S, SM and T) lead to extremely strong rejections, so the
inflated empirical size is not a problem. Figure~\ref{fsize} also
shows that the Monte Carlo approximation is useful for $N=32$, this is
the sample size we must use in Section~\ref{sapp-te}. The size of
test T is practically indistinguishable from that of test SM.
Figure~\ref{fPower} shows the power of method SM; power curves
for method T are practically the same, method S has higher power.
The simulation study shows that a strong rejection when the test
is applied to real data can be viewed as reliable evidence of
dependence.

%s8 ###
\section{Application to critical
ionospheric frequency and
magnetic curves} \label{sapp-te}

In this section we apply the correlation test,
which uses the estimation methodology of Sections
\ref{smean} and \ref{sPC}, to foF2 and magnetic curves.

%f7 ###

\textit{Description of the data.} The F2 layer of the ionosphere is the
upper part of the F layer shown in Figure \ref{fai}. The F2 layer
electron critical frequency, foF2, is measured using an instrument
called the ionosonde, a type of radar. The foF2 frequency is used to
estimate the location of the peak electron density, so an foF2 trend
corresponds to a trend in the average height of the ionosphere over a
spatial location. The foF2 data have therefore been used to test the
hypothesis of ionospheric global cooling discussed in the
\hyperref[si]{Introduction}. Hourly values of foF2 are available from the SPIDR database
\url{http://spidr.ngdc.noaa.gov/spidr/} for more than 200
ionosondes. We use monthly averages for 32 selected ionosondes, with
sufficiently complete records, for the period 1964--1992.
Their locations are shown in Figure~\ref{floc}. Three
typical foF2 curves are shown in Figure \ref{ffoF2}.
We omit the details of the procedure for obtaining curves like those
shown in Figure \ref{ffoF2}, but we emphasize that it requires a great
deal of work. In particular, the SPIDR data suffer from two problems.
First, for some data, the amplitude is artificially magnified
ten times, and needs to be converted into standard units (MHz).
Second, in many cases, missing observations are not replaced by the
standard notation $9999$, but rather just skipped. Thus, if one wants
to use equally-spaced time series, skipped data must be found and
replaced by missing values. For filling in missing values, we perform
linear interpolation. We developed a customized C++ code to handle
these issues. We emphasize that one of the reasons why this global data
set has not been analyzed so far is that useable data have been derived
only over relatively small regions, for example, Western Europe, and
more often
for a single location.

%f8 ###
\begin{figure}[b]

\includegraphics{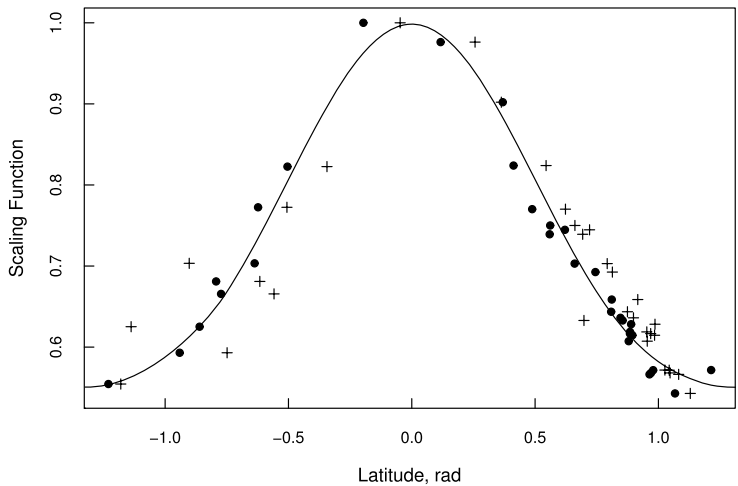}

\caption{Dots represent the scaling function $G_L (\mathbf{s}_i)$ in the magnetic
coordinate system and crosses are the same in the geographic coordinate system.
Line is the best fit for $G_L$ in the magnetic coordinate system.}
\label{fFig2}
\end{figure}

As explained in Section \ref{si}, the foF2 data are used to test
hypotheses on long-term ionospheric trends. We thus removed annual and
higher frequency variations using $16$ month averaging with MODWT
filter; see Chapter 5 of Percival and Walden (\citeyear
{percival-walden-2000}). This leads
to 32 time series at different locations, each containing $336$
equally-spaced temporal observations. The amplitude of the foF2
curves exhibits a nonlinear latitudinal trend; it decreases as the
latitude increases; see Figure~\ref{ffoF2}. To remove this trend,
which may potentially bias the test, we assume that the foF2 signal,
$F(\mathbf{s}; t)$, at location~$\mathbf{s}$ follows the model
%
%e8.1 ###
\begin{equation} \label{eX}
F (\mathbf{s}; t) = G(L(\mathbf{s}))X(\mathbf{s}; t),
\end{equation}
where $X(\mathbf{s}; t)$ is a constant amplitude field,
and $G(\cdot)$ is a scaling function
which depends only on the \textit{magnetic} latitude $L$ (in radians).
Since the trend in the amplitude of $F(\mathbf{s}; t)$ is
caused by the solar
radiation which is nonlinearly proportional to the zenith angle,
we postulate that the function $G(\cdot)$ has the form
%
%e8.2 ###
\begin{equation} \label{eGL}
G(L) = a + b\cos^c (L).
\end{equation}
The parameters $a,b,c$ are estimated as follows.
Let $\mathbf{s}_0$ be the position of the ionosonde closest to the
magnetic equator.
For identifiability, we set $G(L(\mathbf{s}_0)) = 1$.
For the remaining locations $\mathbf{s}_k$, we compute
$\hat G(L(\mathbf{s}_k))$ as the average, over all 336 time points $t_j$
of the
ratio $F(\mathbf{s}_k;t_j)/F(\mathbf{s}_0;t_j)$. Figure~\ref{fFig2}
shows these ratios as a function of the magnetic and geographic
latitude. The ratios in the magnetic latitude show much less
spread, and this is another reason why we work with the magnetic latitude.
The curve $G(L)$ (\ref{eGL}) is fitted to the $\hat G(L(\mathbf{s}_k))$
in magnetic latitude by nonlinear least squares. The fitted values
are $a = 0.5495$, $b = 0.4488$, $c = 4.2631$.\vadjust{\goodbreak}

We now describe how we construct the curves that reflect the relevant
long-term changes in the internal magnetic field of the earth. The
height of the F2 layer (and so the foF2 frequency) can be affected by
a vertical plasma drift which responds to the magnetic field. The
vertical plasma drift is due to the wind effect, and is given by [we
use the same notation as in Mikhailov and Marin (\citeyear
{mikhailov-marin-2001})]
\[
W = (V_{nx} \cos D - V_{ny} \sin D) \sin I \cos I + V_{nz} \sin^2 I.
\]
In the above formula,
$V_{nx}$, $V_{ny}$ and $V_{nz}$ are, respectively, meridional
(parallel to
constant longitude lines), zonal (parallel to constant latitude lines) and
vertical components of the thermospheric neutral wind; $I$ and $D$ are
inclination and declination of the earth magnetic field. Detailed
figures are provided in Chapter 13 of Kivelson and Russell (\citeyear
{kivelson-russell-1997}).
Usually $V_{nz}\ll V_{nx} , V_{ny},$ and assuming that the difference
between magnetic and geographic coordinates,~$D$, is small (at least
for low- and mid-latitude regions), we can simplify the above formula
to $ W = V_{nx} \sin I \cos I. $ Thus, only the meridional thermospheric
wind is significant. Measuring neutral wind components
($V_{nx}$, $V_{ny}$,~$V_{nz}$) is difficult, and
long-term wind records are not available. We therefore replace
$V_{nx}$ by its average. For the test of correlation, the
specific value of this average plays no role, so we define the
magnetic curves as
%
%e8.3 ###
\begin{equation} \label{eY}
Y(\mathbf{s}; t) = \sin I(\mathbf{s};t) \cos I(\mathbf{s};t).
\end{equation}
The curves $ I(\mathbf{s};t)$ are computed using the
international geomagnetic reference field (IGRF); the software
is available at \texttt{\href{http://www.ngdc.noaa.gov/IAGA/vmod/}{http://www.ngdc.noaa.gov/}
\href{http://www.ngdc.noaa.gov/IAGA/vmod/}{IAGA/vmod/}}.

The test is applied to the curves $X(\mathbf{s}_k; t)$ defined by (\ref{eX})
and (\ref{eGL}), and to the curves $Y(\mathbf{s}_k; t)$ defined by (\ref{eY}).

\textit{Application of the correlation test.} We first estimate and
subtract the mean functions of the fields $X(\mathbf{s}_k)$ and $Y(\mathbf{s}_k)$
using method M2 (the other spatial methods give practically the same
estimates). The principal components $v_i$ and $u_i$ are
estimated using method CM2 (method CM3 gives practically the same curves).

%f9 ###
\begin{figure}

\includegraphics{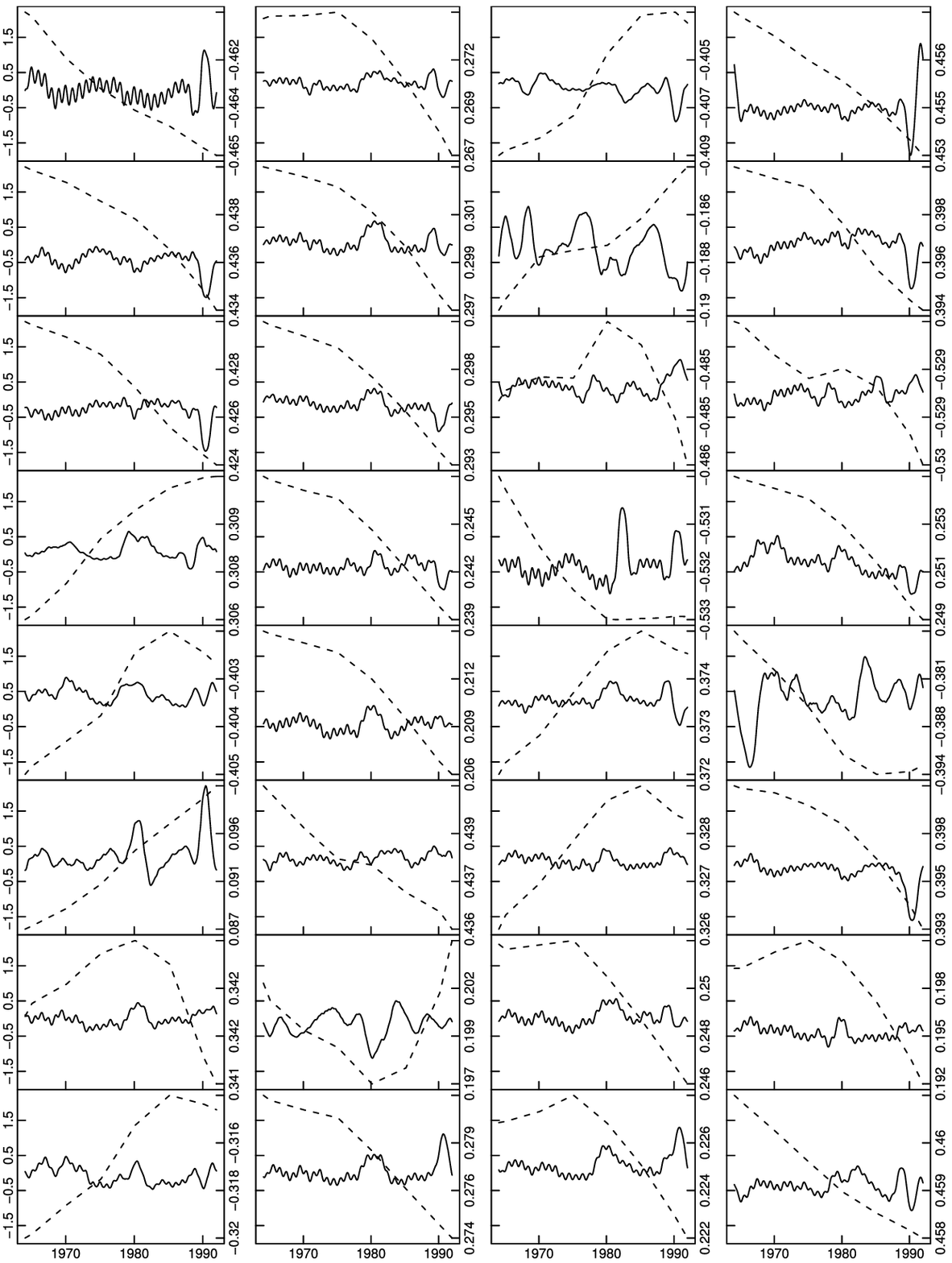}

\caption{Transformed and centered foF2 curves (continuous) and
centered magnetic curves (dashed) at 32 locations denoted with circles
in Figure~\protect\ref{floc}. The scales for the two families of curves are
different. The foF2 curves have the same scale, it is shown on the
right vertical axes in MHz. The scale of the magnetic curves changes,
it is shown on the right vertical axes in each box (unitless).}
\label{fdatascaled}
\end{figure}

%t2 ###
\begin{table}
\tabcolsep=0pt
\caption{$P$-values of the correlation tests applied to the
transformed foF2 data.
The first column shows the number of FPCs,
the second column shows cumulative variances computed as
the ratios of the eigenvalues estimated using method CM2.
Testing procedures S, SM and T are defined in Section \protect\ref{ssp-cor}.
The ``simple'' procedure neglects the spatial dependence of the curves}
\label{tbsimulations}
\begin{tabular*}{\textwidth}{@{\extracolsep{\fill}}lccccc@{}}
\hline
 &  &\multicolumn{3}{c}{\textbf{Spatial}}  \\[-5pt]
 &  &\multicolumn{3}{c}{\hrulefill}\\
$\boldsymbol{p}$& \textbf{CV, $\boldsymbol{\%}$}& \textbf{S} & \textbf{SM} & \textbf{T} & \textbf{Simple}\\
\hline
1 & 47.88 & %22.10 &
$6.22\cdot10^{-5}$\hphantom{$^1$}& $3.05\cdot10^{-4}$ & $3.05\cdot10^{-4}$ &
$0.035$ \\
2 & 62.59 & %25.27 &
$3.26\cdot10^{-6}$\hphantom{$^1$}& $2.91\cdot10^{-4}$ & $2.99\cdot10^{-4}$ &
$0.095$ \\
3 & 73.67 & %37.03 &
$4.53\cdot10^{-8}$\hphantom{$^1$}& $2.43\cdot10^{-4}$ & $2.32\cdot10^{-4}$ &
$0.043$ \\
4 & 84.40 & %127.30 &
$1.47\cdot10^{-26}$& $1.6 \cdot10^{-7}$\hphantom{0} & $2.24 \cdot10^{-5}$ &
$0.039$ \\
5 & 88.70 & %128.50 &
$4.95\cdot10^{-26}$& $2.6 \cdot10^{-7}$\hphantom{0} &$2.27 \cdot10^{-5}$&
$0.046$ \\
6 & 92.21 & %136.08 &
$6.73\cdot10^{-27}$&$5.9 \cdot10^{-7}$\hphantom{0} &$2.21 \cdot10^{-5}$& $0.060$
\\
7 & 94.57 & %165.60 &
$2.12\cdot10^{-32}$&$1.6\cdot10^{-7}$\hphantom{0} &$1.92\cdot10^{-5}$& $0.030$ \\
\hline
\end{tabular*}
\end{table}

We apply the test, for all $1 \le p \le7$ and $q=1$. The first seven
eigenvalues of the field $X$ [computed per (\ref{e5p}) or its analog
for method CM2] explain about 95\% of the variance. The first
eigenvalue of the field $Y$ explains about 99\% of the variance. The
eigenfunction $u_1$ is approximately equal to the linear function:
$u_1(t) \sim t$. This means that at any location, after removing the
average, the magnetic field either linearly increases or decreases,
with slopes depending on the location; see Figure~\ref{fdatascaled}.
To lighten the notation, we drop the ``hats'' from the estimated scores
and denote the zero mean vector
$[xi_i(\mathbf{s}_1),\ldots,\xi_i(\mathbf{s}_N)]^T$ by~$\bxi_i$,
and
$[\eta_1(\mathbf{s}_1),\ldots,\eta_1(\mathbf{s}_N)]^T$ by~$\bfeta$.
The covariances~$\bSig_{\xi_i}$ and~$\bSig_{\eta}$
are estimated using parametric spatial
models determined by the inspection of the empirical variograms.
In this application, it is sufficient to use two covariance models:
%
%e8.4 ###
\begin{eqnarray} \label{eE-G}
\mbox{Gaussian:}\  c(\mathbf{s}_k, \mathbf{s}_\ell)
= c_0 + \sigma^2\exp\{-d^2(k,\ell)/\rho^2\},\nonumber
\\[-8pt]
\\[-8pt]
\mbox{Exponential:}\  c(\mathbf{s}_k, \mathbf{s}_\ell)
= c_0 + \sigma^2\exp\{-d(k,\ell)/\rho\}.
\nonumber
\end{eqnarray}
When the scores do not have a spatial structure,
we use the sample variance (flat variogram).
The estimated models and their parameters are listed in Table~\ref{tbres}.

The $P$-values for different numbers of FPCs $1\leq p \leq7$ are
summarized in Table~\ref{tbsimulations}. Independent of $p$ and
a specific implementation of the test, all
$P$-values are very small, and so the rejection of the null hypothesis
is conclusive; we conclude that there is a statistically significant
correlation between the foF2 curves $X(\mathbf{s}_k)$ and the magnetic
curves $Y(\mathbf{s}_k)$. We also applied a version of our test which neglects
any spatial dependence, this is the test proposed by
Kokoszka et al. (\citeyear{kokoszka-maslova-s-z-2008}). The $P$-values
hover around the
5\% level, but still point toward rejection. The evidence is, however,
much less clear cut. This may partially explain why this issue has
been a matter of much debate in the space physics community.
The correlation between the foF2 and magnetic curves is far from obvious.
Figure~\ref{fdatascaled} shows these pairs at all 32 locations. It is
hard to conclude by eye that the direction of the magnetic field change
impacts the foF2 curves.

\textit{Discussion.}
A very important role in our analysis is played by the transformation
(\ref{eGL}). Applying the test to the original foF2 curves, $F(\mathbf{s}_k;
t)$, gives the $P$-values 0.209 ($p=1$) and 0.011 ($p=2$) for the
spatial S test, and 0.707 ($p=1$), 0.185 ($p=2$), 0.139 ($p=3$) for
the ``simple'' test. As explained above, the amplitude of the field
$F(\mathbf{s}_k; t)$ evolves with the latitude. This invalidates the
assumption of a mean function which is independent of the spatial
location. Thus, even for the spatial test, the mean function confounds
the first FPC. However, the spatial estimation of the mean function
and of the FPCs ``quickly corrects'' for the violation of assumptions, and
the null hypothesis is rejected for $p\ge2$. When the spatial
structure is neglected (and no latitudinal transformation is applied) no
correlation between the foF2 curves and magnetic curves is found.

%f10 ###
\begin{figure}

\includegraphics{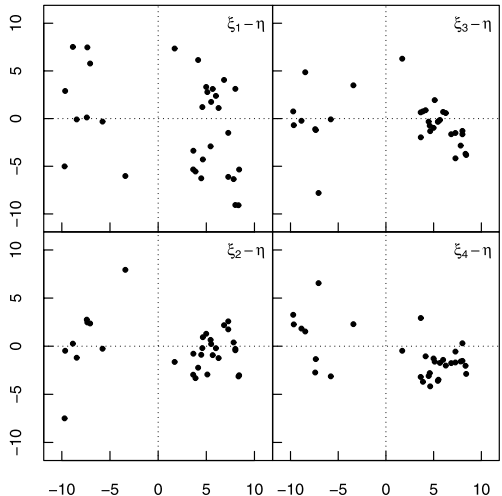}

\caption{Scatter plots of the scores $\bxi_i, i=1,2,3,4$ of the foF2
curves, vertical axes,
against the scores $\bfeta$ of the magnetic curves, horizontal axes.}
\label{fscatter}
\end{figure}

The rejection of the null hypothesis means that after
adjusting the foF2 curves for the latitude and the global mean,
their regional variability is correlated with the regional changes
in the magnetic field. This conclusion agrees with recent
space physics research [see
Cnossen and Richmond (\citeyear{cnossen-richmond-2008}) and
La\v{s}tovi\v{c}ka (\citeyear{lastovicka-2009})],
and can, to some extent, be visually confirmed, post-analysis,
by the examination of the scatter plots shown in Figure~\ref{fscatter}.
It implies that long-term  magnetic trends
must be considered as additional covariates in testing for  long-term
trends in the foF2 curves. The main covariate is the solar activity
which drives the shape of the mean function, but, as explained in
the \hyperref[si]{Introduction}, the impact of the concentration of the greenhouse
gases is of particular interest; see Qian et al. (\citeyear
{qian-burns-solomon-roble-2009}), among many other contributions.

A broader conclusion of the work presented in this paper is that
methods of functional data analysis must be applied with care
to curves obtained at spatial locations. Neglecting the spatial dependence
can lead to incorrect conclusions and biased estimates.
The same applies to space physics research. If trends or models
are estimated separately at each spatial location, one should not
rely on results obtained by some form of a simple averaging. This is,
however, the prevailing approach. Interestingly, the results
related to global ionospheric trends are often on the borderline
of statistical significance. Standard $t$-tests lead
either to rejection or acceptance, depending on a specific method used
(a similar phenomenon is observed in the last column
of Table~\ref{tbsimulations}). It is hoped that the methodology
developed in this paper will be useful in addressing such issues.

\section*{Acknowledgment}
We are grateful to Levan
Lomidze for his help in processing the ionosonde data.

%suskaldyti doi

% imsref loaded by smiklovaite, 2012-01-30 10:13:34
% imsref loaded by smiklovaite, 2012-01-30 10:14:41
% imsref loaded by smiklovaite, 2012-01-30 10:20:13
% imsref loaded by smiklovaite, 2012-01-30 10:22:59

\printaddresses


\begin{thebibliography}{31}
% BibTex style file: ims.bst, 2011-05-30
% Default style options (sort=0,type=number).
% Used options (sort=1,type=nameyear).

%b1 ###
\bibitem[\protect\citeauthoryear{Bel et~al.}{2011}]{bel-2011}
\begin{barticle}[mr]
\bauthor{\bsnm{Bel},~\bfnm{Liliane}\binits{L.}},
  \bauthor{\bsnm{Bar-Hen},~\bfnm{Avner}\binits{A.}},
  \bauthor{\bsnm{Petit},~\bfnm{R{\'e}my}\binits{R.}} \AND
  \bauthor{\bsnm{Cheddadi},~\bfnm{Rachid}\binits{R.}}
(\byear{2011}).
\btitle{Spatio-temporal functional regression on paleoecological data}.
\bjournal{J. Appl. Stat.}
\bvolume{38}
\bpages{695--704}.
\bid{doi={10.1080/02664760903563650}, issn={0266-4763}, mr={2773575}}
\bptok{imsref}%
\end{barticle}
\endbibitem

%b2 ###
\bibitem[\protect\citeauthoryear{Cnossen and
  Richmond}{2008}]{cnossen-richmond-2008}
\begin{barticle}[auto:STB|2012/01/27|08:30:54]
\bauthor{\bsnm{Cnossen},~\bfnm{I.}\binits{I.}} \AND
  \bauthor{\bsnm{Richmond},~\bfnm{A.~D.}\binits{A.~D.}}
(\byear{2008}).
\btitle{Modelling the effects of changes in the Earth's magnetic field from
  1957 to 1997 on the ionospheric hmF2 and foF2 parameters}.
\bjournal{Journal of Atmospheric and Solar-Terrestrial Physics}
\bvolume{70}
\bpages{1512--1524}.
\bptok{imsref}%
\end{barticle}
\endbibitem

%b3 ###
\bibitem[\protect\citeauthoryear{Delicado et~al.}{2010}]{delicado-2010}
\begin{barticle}[mr]
\bauthor{\bsnm{Delicado},~\bfnm{P.}\binits{P.}},
  \bauthor{\bsnm{Giraldo},~\bfnm{R.}\binits{R.}},
  \bauthor{\bsnm{Comas},~\bfnm{C.}\binits{C.}} \AND
  \bauthor{\bsnm{Mateu},~\bfnm{J.}\binits{J.}}
(\byear{2010}).
\btitle{Statistics for spatial functional data: Some recent contributions}.
\bjournal{Environmetrics}
\bvolume{21}
\bpages{224--239}.
\bid{issn={1180-4009}, mr={2842240}}
\bptok{imsref}%
\end{barticle}
\endbibitem

%b4 ###
\bibitem[\protect\citeauthoryear{Finkenst{\"a}dt, Held and
  Isham}{2007}]{finkenstaedt-etal-2007}
\begin{bbook}[mr]
\beditor{\bsnm{Finkenst{\"a}dt},~\bfnm{B{\"a}rbel}\binits{B.}},
  \beditor{\bsnm{Held},~\bfnm{Leonhard}\binits{L.}} \AND
  \beditor{\bsnm{Isham},~\bfnm{Valerie}\binits{V.}}, eds.
(\byear{2007}).
\btitle{Statistical Methods for Spatio-Temporal Systems}.
\bseries{Monographs on Statistics and Applied Probability}
\bvolume{107}.
\bpublisher{Chapman \& Hall/CRC}, \baddress{Boca Raton, FL}.
\bid{mr={2307967}}
\bptok{imsref}%
\end{bbook}
\endbibitem






%b6 ###
\bibitem[\protect\citeauthoryear{Gelfand et~al.}{2010}]{gelfand-etal-2010}
\begin{bbook}[mr]
\beditor{\bsnm{Gelfand},~\bfnm{Alan~E.}\binits{A.~E.}},
  \beditor{\bsnm{Diggle},~\bfnm{Peter~J.}\binits{P.~J.}},
  \beditor{\bsnm{Fuentes},~\bfnm{Montserrat}\binits{M.}} \AND
  \beditor{\bsnm{Guttorp},~\bfnm{Peter}\binits{P.}}, eds.
(\byear{2010}).
\btitle{Handbook of Spatial Statistics}.
\bpublisher{CRC Press}, \baddress{Boca Raton, FL}.
\bid{mr={2761512}}
\bptok{imsref}%
\end{bbook}
\endbibitem

%b7 ###
\bibitem[\protect\citeauthoryear{Giraldo, Delicado and
  Mateu}{2011a}]{giraldo-delicado-mateu-2011a}
\begin{barticle}[auto:STB|2012/01/27|08:30:54]
\bauthor{\bsnm{Giraldo},~\bfnm{R.}\binits{R.}},
  \bauthor{\bsnm{Delicado},~\bfnm{P.}\binits{P.}} \AND
  \bauthor{\bsnm{Mateu},~\bfnm{J.}\binits{J.}}
(\byear{2011}a).
\btitle{Ordinary kriging for function-valued spatial data}.
\bjournal{Environ. Ecol. Stat.}
\bvolume{18}
\bpages{411--426}.
\bptok{imsref}%
\end{barticle}
\endbibitem

%b8 ###
\bibitem[\protect\citeauthoryear{Giraldo, Delicado and
  Mateu}{2011b}]{giraldo-delicado-mateu-2011b}
\begin{bmisc}[auto:STB|2012/01/27|08:30:54]
\bauthor{\bsnm{Giraldo},~\bfnm{R.}\binits{R.}},
  \bauthor{\bsnm{Delicado},~\bfnm{P.}\binits{P.}} \AND
  \bauthor{\bsnm{Mateu},~\bfnm{J.}\binits{J.}}
(\byear{2011}b).
\bhowpublished{A generalization of cokriging and multivariable spatial
  prediction for functional data. Technical report, Univ. Polit\'ecnica
  de Catalunya, Barcelona}.
\bptok{imsref}%
\end{bmisc}
\endbibitem

%b9 ###
\bibitem[\protect\citeauthoryear{Gneiting}{2011}]{gneiting-2011}
\begin{barticle}[auto:STB|2012/01/27|08:30:54]
\bauthor{\bsnm{Gneiting},~\bfnm{T.}\binits{T.}}
(\byear{2011}).
\btitle{Making and evaluating point forecasts}.
\bjournal{J. Amer. Statist. Assoc.}
\bvolume{106}
\bpages{746--762}.
\bptok{imsref}%
\end{barticle}
\endbibitem

%b10 ###
\bibitem[\protect\citeauthoryear{H{\"o}rmann and
  Kokoszka}{2012}]{hormann-kokoszka-2011}
\begin{bmisc}[auto:STB|2012/01/27|08:30:54]
\bauthor{\bsnm{H{\"o}rmann},~\bfnm{S.}\binits{S.}} \AND
  \bauthor{\bsnm{Kokoszka},~\bfnm{P.}\binits{P.}}
(\byear{2012}).
\bhowpublished{Consistency of the mean and the principal components of
  spatially distributed functional data. \textit{Bernoulli}. To appear}.
\bptok{imsref}%
\end{bmisc}
\endbibitem

%b11 ###
\bibitem[\protect\citeauthoryear{Horv{\'a}th and Kokoszka}{2012}]{HKbook}
\begin{bbook}[auto:STB|2012/01/27|08:30:54]
\bauthor{\bsnm{Horv{\'a}th},~\bfnm{L.}\binits{L.}} \AND
  \bauthor{\bsnm{Kokoszka},~\bfnm{P.}\binits{P.}}
(\byear{2012}).
\btitle{Inference for Functional Data with Applications}.
%Springer Series in Statistics.}
\bpublisher{Springer}, \baddress{Berlin}.
\bptok{imsref}%
\end{bbook}
\endbibitem

%b5 ###
\bibitem[\protect\citeauthoryear{Kivelson and Russell}{1997}]{kivelson-russell-1997}
\begin{bbook}[mr]
\beditor{\bsnm{Kivelson},~\bfnm{M. G.}\binits{M. G.}} \AND
  \beditor{\bsnm{Russell},~\bfnm{C. T.}\binits{C. T.}}, eds.
(\byear{1997}).
\btitle{Introduction to Space Physics}.
\bpublisher{Cambridge Univ. Press}, \baddress{Cambridge}.
\bptok{imsref}%
\end{bbook}
\endbibitem

%b12 ###
\bibitem[\protect\citeauthoryear{Kokoszka
  et~al.}{2008}]{kokoszka-maslova-s-z-2008}
\begin{barticle}[mr]
\bauthor{\bsnm{Kokoszka},~\bfnm{Piotr}\binits{P.}},
  \bauthor{\bsnm{Maslova},~\bfnm{Inga}\binits{I.}},
  \bauthor{\bsnm{Sojka},~\bfnm{Jan}\binits{J.}} \AND
  \bauthor{\bsnm{Zhu},~\bfnm{Lie}\binits{L.}}
(\byear{2008}).
\btitle{Testing for lack of dependence in the functional linear model}.
\bjournal{Canad. J. Statist.}
\bvolume{36}
\bpages{207--222}.
\bid{doi={10.1002/cjs.5550360203}, issn={0319-5724}, mr={2431682}}
\bptok{imsref}%
\end{barticle}
\endbibitem

%b13 ###
\bibitem[\protect\citeauthoryear{La{\v{s}}tovi{\v{c}}ka}{2009}]{lastovicka-2009}
\begin{barticle}[auto:STB|2012/01/27|08:30:54]
\bauthor{\bsnm{La{\v{s}}tovi{\v{c}}ka},~\bfnm{J.}\binits{J.}}
(\byear{2009}).
\btitle{Global pattern of trends in the upper atmosphere and ionosphere: Recent
  progress. \textit{Journal of Atmospheric and Solar-Terrestrial}}.
\bjournal{Physics}
\bvolume{71}
\bpages{1514--1528}.
\bptok{imsref}%
\end{barticle}
\endbibitem

%b14 ###
\bibitem[\protect\citeauthoryear{La{\v{s}}tovi{\v{c}}ka
  et~al.}{2008}]{lastovicka-2008}
\begin{barticle}[auto:STB|2012/01/27|08:30:54]
\bauthor{\bsnm{La{\v{s}}tovi{\v{c}}ka},~\bfnm{J.}\binits{J.}},
  \bauthor{\bsnm{Akmaev},~\bfnm{R.~A.}\binits{R.~A.}},
  \bauthor{\bsnm{Beig},~\bfnm{G.}\binits{G.}},
  \bauthor{\bsnm{Bremer},~\bfnm{J.}\binits{J.}},
  \bauthor{\bsnm{Emmert},~\bfnm{J.~T.}\binits{J.~T.}},
  \bauthor{\bsnm{Jacobi},~\bfnm{C.}\binits{C.}},
  \bauthor{\bsnm{Jarvis},~\bfnm{J.~M.}\binits{J.~M.}},
  \bauthor{\bsnm{Nedoluha},~\bfnm{G.}\binits{G.}},
  \bauthor{\bsnm{Portnyagin},~\bfnm{Yu.~I.}\binits{Y.~I.}} \AND
  \bauthor{\bsnm{Ulich},~\bfnm{T.}\binits{T.}}
(\byear{2008}).
\btitle{Emerging pattern of global change in the upper atmosphere and
  ionosphere}.
\bjournal{Annales Geophysicae}
\bvolume{26}
\bpages{1255--1268}.
\bptok{imsref}%
\end{barticle}
\endbibitem

%b15 ###
\bibitem[\protect\citeauthoryear{Maslova
  et~al.}{2009}]{maslova-kokoszka-s-z-2009}
\begin{barticle}[auto:STB|2012/01/27|08:30:54]
\bauthor{\bsnm{Maslova},~\bfnm{I.}\binits{I.}},
  \bauthor{\bsnm{Kokoszka},~\bfnm{P.}\binits{P.}},
  \bauthor{\bsnm{Sojka},~\bfnm{J.}\binits{J.}} \AND
  \bauthor{\bsnm{Zhu},~\bfnm{L.}\binits{L.}}
(\byear{2009}).
\btitle{Removal of nonconstant daily variation by means of wavelet and
  functional data analysis}.
\bjournal{Journal of Geophysical Research}
\bvolume{114}
\bpages{A03202}.
\bptok{imsref}%
\end{barticle}
\endbibitem

%b16 ###
\bibitem[\protect\citeauthoryear{Maslova
  et~al.}{2010a}]{maslova-kokoszka-s-z-2010}
\begin{barticle}[auto:STB|2012/01/27|08:30:54]
\bauthor{\bsnm{Maslova},~\bfnm{I.}\binits{I.}},
  \bauthor{\bsnm{Kokoszka},~\bfnm{P.}\binits{P.}},
  \bauthor{\bsnm{Sojka},~\bfnm{J.}\binits{J.}} \AND
  \bauthor{\bsnm{Zhu},~\bfnm{L.}\binits{L.}}
(\byear{2010}a).
\btitle{Estimation of Sq variation by means of multiresolution and principal
  component analyses}.
\bjournal{Journal of Atmospheric and Solar-Terrestial Physics}
\bvolume{72}
\bpages{625--632}.
\bptok{imsref}%
\end{barticle}
\endbibitem

%b17 ###
\bibitem[\protect\citeauthoryear{Maslova
  et~al.}{2010b}]{maslova-kokoszka-s-z-2010PSS}
\begin{barticle}[auto:STB|2012/01/27|08:30:54]
\bauthor{\bsnm{Maslova},~\bfnm{I.}\binits{I.}},
  \bauthor{\bsnm{Kokoszka},~\bfnm{P.}\binits{P.}},
  \bauthor{\bsnm{Sojka},~\bfnm{J.}\binits{J.}} \AND
  \bauthor{\bsnm{Zhu},~\bfnm{L.}\binits{L.}}
(\byear{2010}b).
\btitle{Statistical significance testing for the association of magnetometer
  records at high-, mid- and low latitudes during substorm days}.
\bjournal{Planetary and Space Science}
\bvolume{58}
\bpages{437--445}.
\bptok{imsref}%
\end{barticle}
\endbibitem

%b18 ###
\bibitem[\protect\citeauthoryear{Mikhailov and
  Marin}{2001}]{mikhailov-marin-2001}
\begin{barticle}[auto:STB|2012/01/27|08:30:54]
\bauthor{\bsnm{Mikhailov},~\bfnm{A.~V.}\binits{A.~V.}} \AND
  \bauthor{\bsnm{Marin},~\bfnm{D.}\binits{D.}}
(\byear{2001}).
\btitle{An interpretation of the foF2 and hmF2 long-term trends in the
  framework of the geomagnetic control concept}.
\bjournal{Annales Geophysicae}
\bvolume{19}
\bpages{733--748}.
\bptok{imsref}%
\end{barticle}
\endbibitem

%b19 ###
\bibitem[\protect\citeauthoryear{Nerini, Monestiez and
  Mant{\'e}}{2010}]{nerini-monestiez-mantea-2010}
\begin{barticle}[mr]
\bauthor{\bsnm{Nerini},~\bfnm{David}\binits{D.}},
  \bauthor{\bsnm{Monestiez},~\bfnm{Pascal}\binits{P.}} \AND
  \bauthor{\bsnm{Mant{\'e}},~\bfnm{Claude}\binits{C.}}
(\byear{2010}).
\btitle{Cokriging for spatial functional data}.
\bjournal{J. Multivariate Anal.}
\bvolume{101}
\bpages{409--418}.
\bid{doi={10.1016/j.jmva.2009.03.005}, issn={0047-259X}, mr={2564350}}
\bptok{imsref}%
\end{barticle}
\endbibitem

%b20 ###
\bibitem[\protect\citeauthoryear{Percival and
  Walden}{2000}]{percival-walden-2000}
\begin{bbook}[mr]
\bauthor{\bsnm{Percival},~\bfnm{Donald~B.}\binits{D.~B.}} \AND
  \bauthor{\bsnm{Walden},~\bfnm{Andrew~T.}\binits{A.~T.}}
(\byear{2000}).
\btitle{Wavelet Methods for Time Series Analysis}.
\bseries{Cambridge Series in Statistical and Probabilistic Mathematics}
\bvolume{4}.
\bpublisher{Cambridge Univ. Press}, \baddress{Cambridge}.
\bid{mr={1770693}}
\bptok{imsref}%
\end{bbook}
\endbibitem

%b21 ###
\bibitem[\protect\citeauthoryear{Qian
  et~al.}{2009}]{qian-burns-solomon-roble-2009}
\begin{barticle}[auto:STB|2012/01/27|08:30:54]
\bauthor{\bsnm{Qian},~\bfnm{L.}\binits{L.}},
  \bauthor{\bsnm{Burns},~\bfnm{A.~G.}\binits{A.~G.}},
  \bauthor{\bsnm{Solomon},~\bfnm{S.~C.}\binits{S.~C.}} \AND
  \bauthor{\bsnm{Roble},~\bfnm{R.~G.}\binits{R.~G.}}
(\byear{2009}).
\btitle{The effect of carbon dioxide cooling on trends in the F2-layer
  ionosphere}.
\bjournal{Journal of Atmospheric and Solar-Terrestrial Physics}
\bvolume{71}
\bpages{1592--1601}.
\bptok{imsref}%
\end{barticle}
\endbibitem

%b22 ###
\bibitem[\protect\citeauthoryear{Ramsay, Hooker and
  Graves}{2009}]{ramsay-hooker-graves-2009}
\begin{bbook}[auto:STB|2012/01/27|08:30:54]
\bauthor{\bsnm{Ramsay},~\bfnm{J.}\binits{J.}},
  \bauthor{\bsnm{Hooker},~\bfnm{G.}\binits{G.}} \AND
  \bauthor{\bsnm{Graves},~\bfnm{S.}\binits{S.}}
(\byear{2009}).
\btitle{Functional Data Analysis with R and MATLAB}.
\bpublisher{Springer}, \baddress{New York}.
\bptok{imsref}%
\end{bbook}
\endbibitem

%b23 ###
\bibitem[\protect\citeauthoryear{Rishbeth}{1990}]{rishbeth-1990}
\begin{barticle}[auto:STB|2012/01/27|08:30:54]
\bauthor{\bsnm{Rishbeth},~\bfnm{H.}\binits{H.}}
(\byear{1990}).
\btitle{A greenhouse effect in the ionosphere?}
\bjournal{Planetary and Space Science}
\bvolume{38}
\bpages{945--948}.
\bptok{imsref}%
\end{barticle}
\endbibitem

%b24 ###
\bibitem[\protect\citeauthoryear{Roble and
  Dickinson}{1989}]{roble-dickinson-1989}
\begin{barticle}[auto:STB|2012/01/27|08:30:54]
\bauthor{\bsnm{Roble},~\bfnm{R.~G.}\binits{R.~G.}} \AND
  \bauthor{\bsnm{Dickinson},~\bfnm{R.~E.}\binits{R.~E.}}
(\byear{1989}).
\btitle{How will changes in carbon dioxide and methane modify the mean
  structure of the mesosphere and thermosphere?}
\bjournal{Geophysical Research Letters}
\bvolume{16}
\bpages{1441--1444}.
\bptok{imsref}%
\end{barticle}
\endbibitem

%b25 ###
\bibitem[\protect\citeauthoryear{Schabenberger and
  Gotway}{2005}]{schabenberger-gotway-2005}
\begin{bbook}[mr]
\bauthor{\bsnm{Schabenberger},~\bfnm{Oliver}\binits{O.}} \AND
  \bauthor{\bsnm{Gotway},~\bfnm{Carol~A.}\binits{C.~A.}}
(\byear{2005}).
\btitle{Statistical Methods for Spatial Data Analysis}.
\bpublisher{Chapman \& Hall/CRC}, \baddress{Boca Raton, FL}.
\bid{mr={2134116}}
\bptok{imsref}%
\end{bbook}
\endbibitem

%b26 ###
\bibitem[\protect\citeauthoryear{Sz{\'e}kely, Rizzo and
  Bakirov}{2007}]{szekely-rizzo-bakirov-2007}
\begin{barticle}[mr]
\bauthor{\bsnm{Sz{\'e}kely},~\bfnm{G{\'a}bor~J.}\binits{G.~J.}},
  \bauthor{\bsnm{Rizzo},~\bfnm{Maria~L.}\binits{M.~L.}} \AND
  \bauthor{\bsnm{Bakirov},~\bfnm{Nail~K.}\binits{N.~K.}}
(\byear{2007}).
\btitle{Measuring and testing dependence by correlation of distances}.
\bjournal{Ann. Statist.}
\bvolume{35}
\bpages{2769--2794}.
\bid{doi={10.1214/009053607000000505}, issn={0090-5364}, mr={2382665}}
\bptok{imsref}%
\end{barticle}
\endbibitem

%b27 ###
\bibitem[\protect\citeauthoryear{Sz{\'e}kely and
  Rizzo}{2009}]{szekely-rizzo-2009}
\begin{barticle}[mr]
\bauthor{\bsnm{Sz{\'e}kely},~\bfnm{G{\'a}bor~J.}\binits{G.~J.}} \AND
  \bauthor{\bsnm{Rizzo},~\bfnm{Maria~L.}\binits{M.~L.}}
(\byear{2009}).
\btitle{Brownian distance covariance}.
\bjournal{Ann. Appl. Stat.}
\bvolume{3}
\bpages{1236--1265}.
\bid{doi={10.1214/09-AOAS312}, issn={1932-6157}, mr={2752127}}
\bptok{imsref}%
\end{barticle}
\endbibitem

%b28 ###
\bibitem[\protect\citeauthoryear{Ulich, Clilverd and
  Rishbeth}{2003}]{ulich-clilverd-risbeth-2003}
\begin{barticle}[auto:STB|2012/01/27|08:30:54]
\bauthor{\bsnm{Ulich},~\bfnm{T.}\binits{T.}},
  \bauthor{\bsnm{Clilverd},~\bfnm{M.~A.}\binits{M.~A.}} \AND
  \bauthor{\bsnm{Rishbeth},~\bfnm{H.}\binits{H.}}
(\byear{2003}).
\btitle{Determining long-term change in the ionosphere}.
\bjournal{Eos, Transactions American Geophysical Union}
\bvolume{84}
\bpages{581--585}.
\bptok{imsref}%
\end{barticle}
\endbibitem

%b29 ###
\bibitem[\protect\citeauthoryear{Wackernagel}{2003}]{wackernagel-2003}
\begin{bbook}[auto:STB|2012/01/27|08:30:54]
\bauthor{\bsnm{Wackernagel},~\bfnm{H.}\binits{H.}}
(\byear{2003}).
\btitle{Multivariate Geostatistics}, \bedition{3rd} ed.
\bpublisher{Springer}, \baddress{New York}.
\bptok{imsref}%
\end{bbook}
\endbibitem

%b30 ###
\bibitem[\protect\citeauthoryear{Yamanishi and
  Tanaka}{2003}]{yamanishi-tanaka-2003}
\begin{barticle}[mr]
\bauthor{\bsnm{Yamanishi},~\bfnm{Yoshihiro}\binits{Y.}} \AND
  \bauthor{\bsnm{Tanaka},~\bfnm{Yutaka}\binits{Y.}}
(\byear{2003}).
\btitle{Geographically weighted functional multiple regression analysis: A
  numerical investigation}.
\bjournal{J. Japanese Soc. Comput. Statist.}
\bvolume{15}
\bpages{307--317}.
%  (Osaka, 2001)}.
\bid{issn={0915-2350}, mr={2027947}}
\bptok{imsref}%
\end{barticle}
\endbibitem

\end{thebibliography}
\end{document}